\documentclass[aip, jcp, longbibliography, twocolumn, floatfix, superscriptaddress, reprint, 10pt, rmp]{revtex4-1}

\usepackage[english]{babel}
\usepackage[utf8]{inputenc}
\usepackage[paperwidth=210mm,paperheight=297mm,centering,hmargin=2cm,vmargin=2.5cm]{geometry}
\usepackage[utf8]{inputenc}
\usepackage{verbatim}
\usepackage{comment}
\usepackage{bm}
\usepackage{amsmath}
\usepackage{amsfonts}
\usepackage{natbib}
\usepackage{pdfpages}
\usepackage{graphics}
\usepackage{float}
\usepackage{booktabs}
\usepackage{hyperref}
\usepackage[capitalize]{cleveref}
\usepackage{multirow}
\usepackage{subcaption}
\usepackage{booktabs}
\usepackage{makecell}
\usepackage{outlines}
\usepackage{lipsum}
\usepackage{siunitx}
\usepackage[version=4]{mhchem}
\usepackage{pdfpages}
\usepackage{pgffor}
\usepackage{xcolor}
\usepackage{ulem}
\usepackage{color,soul}
\ifdefined\DIRACREP
\else
\newcommand{\DIRACREP}{}

\usepackage{physics}
\usepackage{xparse}
\ifdefined\COSMOMATHS
\else
\newcommand{\COSMOMATHS}{}

\newcommand{\mbf}[1]{\ensuremath{\mathbf{#1}}}

\fi %

\NewDocumentCommand{\rep}{s d<| d|>}{%
\IfBooleanTF{#1}{
   \IfValueTF{#2}{
       \IfValueTF{#3}{\braket{#2}{#3}}{\bra{#2}}
       }{
       \IfValueTF{#3}{\ket{#3}}{}
       }
   }{
   \IfValueTF{#2}{
       \IfValueTF{#3}{\braket*{#2}{#3}}{\bra*{#2}}
       }{
       \IfValueTF{#3}{\ket*{#3}}{}
       }
   }
}

\NewDocumentCommand{\rbra}{sm}{\IfBooleanTF{#1}{\rep*<#2|}{\rep<#2|}}
\NewDocumentCommand{\rket}{sm}{\IfBooleanTF{#1}{\rep*|#2>}{\rep|#2>}}
\NewDocumentCommand{\rbraket}{smom}{
    \IfBooleanTF{#1}{
        \IfNoValueTF{#3}{\rep*<#2||#4>}{\rep*<#2|#3\rep*|#4>}
    }{
        \IfNoValueTF{#3}{\rep<#2||#4>}{\rep<#2|#3\rep|#4>}
    }
}

\NewDocumentCommand{\field}{o m e{_} e{^} o e{_} e{^}}{
\IfValueTF{#5}{\overline{
  #2\IfValueT{#3}{_#3}\IfValueT{#4}{^{\otimes #4}} %
  \otimes
  #5\IfValueT{#6}{_#6}\IfValueT{#7}{^{\otimes #7}} %
  \IfValueT{#1}{;#1}
}}{
  \IfValueTF{#4}{\overline{
     #2\IfValueT{#3}{_#3}\IfValueT{#4}{^{\otimes #4}}
     \IfValueT{#1}{;#1}
  }}
  {#2\IfValueT{#3}{_#3}}
}
}

\NewDocumentCommand{\frho}{o e{_} e{^}}{
\field[#1]{\rho}_{#2}^{#3}
}

\newcommand{\bx}{\mbf{x}}

\newcommand{\e}{a}  %

\NewDocumentCommand{\ex}{e_}{
\IfValueTF{#1}{\e_{#1}\bx_{#1}}{\e\bx}
}  %

\NewDocumentCommand{\lm}{e_}{
\IfValueTF{#1}{l_{#1}m_{#1}}{lm}
}
\NewDocumentCommand{\nlm}{e_}{
\IfValueTF{#1}{n_{#1}\lm_{#1}}{n\lm}
}
\NewDocumentCommand{\enlm}{e_}{
\IfValueTF{#1}{\e_{#1}\nlm_{#1}}{\e\nlm}
}
\NewDocumentCommand{\en}{e_}{
\IfValueTF{#1}{\e_{#1}n_{#1}}{\e n}
}
\NewDocumentCommand{\nlk}{e_}{
\IfValueTF{#1}{n_{#1}l_{#1}k_{#1}}{nlk}
}
\NewDocumentCommand{\enlk}{e_}{
\IfValueTF{#1}{\e_{#1}\nlk_{#1}}{\e\nlk}
}
\NewDocumentCommand{\enl}{e_}{
\IfValueTF{#1}{\en_{#1}l_#1}{\en l}
}

\NewDocumentCommand{\nnl}{s}{
\IfBooleanTF{#1}{n_1 n_2 l}{n_1; n_2; l}
}
\NewDocumentCommand{\ennl}{s}{
\IfBooleanTF{#1}{\en_1 \en_2 l}{\en_1; \en_2; l}
}

\NewDocumentCommand{\gslm}{s}{
\IfBooleanTF{#1}{\sigma\lambda\mu}{\sigma;\lambda\mu}
}

\fi %

\ifdefined\COSMOMODELS
\else
\newcommand{\COSMOMODELS}{}
\usepackage{bm,upgreek}

\newcommand{\w}{w}   %

\fi %

\newcommand{\am}[1]{{\color{orange}{#1}}}

\newcommand{\todo}[1]{#1}

\newcommand{\SM}{Supporting Materials}

\newcommand{\rev}[1]{#1}

\makeatletter
\AtBeginDocument{\let\LS@rot\@undefined}
\makeatother

\begin{document}

\setcitestyle{super}

\title{Surface segregation in high-entropy alloys from alchemical machine learning}
\author{Arslan Mazitov}
\affiliation{Laboratory of Computational Science and Modeling, Institute of Materials, \'Ecole Polytechnique F\'ed\'erale de Lausanne, 1015 Lausanne, Switzerland}

\author{Maximilian A.~Springer}
\affiliation{BASF SE, Carl-Bosch-Stra{\ss}e 38, 67056 Ludwigshafen, Germany}

\author{Nataliya Lopanitsyna}
\author{Guillaume Fraux}
\affiliation{Laboratory of Computational Science and Modeling, Institute of Materials, \'Ecole Polytechnique F\'ed\'erale de Lausanne, 1015 Lausanne, Switzerland}

\author{Sandip De}
\affiliation{BASF SE, Carl-Bosch-Stra{\ss}e 38, 67056 Ludwigshafen, Germany}
\author{Michele Ceriotti}
\affiliation{Laboratory of Computational Science and Modeling, Institute of Materials, \'Ecole Polytechnique F\'ed\'erale de Lausanne, 1015 Lausanne, Switzerland}

\onecolumngrid

\begin{abstract}

High-entropy alloys (HEAs), containing several metallic elements in near-equimolar proportions, have long been of interest for their unique mechanical properties. 
More recently, they have emerged as a promising platform for the development of novel heterogeneous catalysts, because of the large design space, and the synergistic effects between their components.
In this work we use a machine-learning potential that can model simultaneously up to 25 transition metals to study the tendency of different elements to segregate at the surface of a HEA.
\rev{We use as a starting point a potential that was previously developed using exclusively crystalline bulk phases, and show that, thanks to the physically-inspired functional form of the model, adding a much smaller number of defective configurations makes it capable of describing surface phenomena. }
We then present several computational studies of surface segregation, including both a simulation of a 25-element alloy, that provides a rough estimate of the relative surface propensity of the various elements, and targeted studies of CoCrFeMnNi  and IrFeCoNiCu, which provide further validation of the model, and insights to guide the modeling and design of alloys for heterogeneous catalysis. 

\end{abstract}

\twocolumngrid

\maketitle

\section{Introduction}

Catalysts are widely used in modern industrial chemistry processes to lower the barriers and thus enhance the rates of a multitude of diverse chemical reactions. 
Among the many different classes of catalysts, a lot of attention has been recently devoted to high-entropy alloys (HEAs). 
Initially introduced by Yeh \cite{yeh+04aem}  and Cantor \cite{cant+04msea} for metallurgic and mechanical applications, HEAs were shown to exhibit promising catalytic \cite{Sun2021,Wang2021, Katiyar2021, Xin2020, Yu2022} and especially electrocatalytic \cite{Huo2021, Zhang2021,Huo2022} behavior. 
The range of HEAs applications for catalysis includes decomposition of water for hydrogen production \cite{Zhang2018, Bondesgaard2019, Glasscott2019, Jin2019, Lacey2019, Liu2019, Qiu2019, Qiu2019a, Gao2020, Huang2020, Wu2020, Katiyar2020}, oxygen reduction \cite{Lacey2019, Qiu2019a, Chen2015, Lffler2018, Li2020, Pedersen2021}, methanol oxidation \cite{Chen2015, Barranco2008, Tsai2009, Wang2014, Yusenko2017, Katiyar2020}, reduction of CO$_2$ and CO molecules \cite{Pedersen2020, Katiyar2020}, and decomposition of ammonia \cite{Zhang2021}. 
The peculiar properties of HEAs are usually attributed to their multicomponent nature. It leads to lattice distortion\cite{Yeh2006} and \textit{sluggish diffusion} \cite{Wang2020} effects, which kinetically stabilize the alloy. Additionally, the ``\textit{cocktail effect}'' \cite{Yeh2006, Yeh2013,Pickering2016, Miracle2017} associated with the synergy between different elements, causes their mechanical and chemical behavior including their enhanced catalytic activity. 

The computational study and modeling of HEAs, and in particular their catalytic properties, is a promising approach to rapidly explore the enormous composition space. However, this is a challenging endeavor: 
Disordered alloys typically require large unit cells to obtain a statistically representative structure, what makes the \textit{ab initio} simulations of HEAs computationally expensive and time consuming. 
\rev{Simulations of both elemental metals and HEAs based on empirical interatomic potentials are much faster, but are usually less accurate\cite{szla+14prb,zuo+20jpcl, lopa+21prm}, especially when used to model multicomponent structures \cite{Ferrari2020a}.
Furthermore, most recent examples of traditional potentials for HEAs focus on a very narrow range of compositions and are specifically optimized for a precise scientific question\cite{fark-caro18jmr,dara+22cms}.
}
Even machine learning interatomic potentials (MLIPs), which typically address this trade-off by approximating the outcome of electronic structure calculations \cite{Farkas2018, Byggmstar2021, rose+21npjcm, zhou+22prb}, cannot be applied straightforwardly. 
Many popular models based on atom-centered descriptors\cite{behl-parr07prl,bart+10prl} suffer from  an exponential scaling of memory and computational requirements with respect to the number of distinct elements. 
Recent developments in Graph Neural Network (GNN) \cite{chen+19cm, Chen2022} and equivariant  \cite{bazt+22ncomm, bata+22nips} models use a chemical space embedding approach that doesn't scale with the size of the chemical space, but lacks interpretability. 
Both approaches also suffer from the exponential increase in data requirements associated with the large composition space.

In our previous work \cite{lopa+23prm}, we introduced a general-purpose machine-learning (ML) model for the study of HEAs with up to 25 transition metals in the composition (the HEA25-4-NN), as well as proposed an efficient approach to generate a training set for it. 
Based on the idea of a physically-motivated and interpretable \textit{alchemical contraction} of the feature space \cite{will+18pccp}, the model demonstrated very promising accuracy, robustness and transferability in terms of various bulk HEA simulation scenarios, which is even more remarkable in light of the difficulty in modeling even \emph{pure} transition metals using MLIPs\cite{owen2023complexity}.
However, catalysis research requires the ability to model surfaces and defective structures, which is beyond the capabilities of a model that is trained exclusively on distorted \emph{fcc} and \emph{bcc} bulk configurations. 
Here we extend the scope of the model to study structure and stability of the surfaces of HEAs, with a particular focus on the problem of surface segregation, which is of great importance for catalysis. 
\rev{
We demonstrate that by extending the bulk training set with a small number (less than 20\%{}) of surface and molten structures, it is possible to achieve a similar level of accuracy also for these defective configurations.
We trace the transferability of the model to the feature space contraction matrix, which is optimized based on the bulk structures and reduces the chemical complexity of the training problem, as well as the data requirements.  
}
We then perform a simulation of the segregation process in a Cantor-style alloy with an equimolar content of all the 25 transition metals included in the training set to evaluate the segregation propensity of different elements. Finally, we study the segregation process in two selected HEAs. 

\section{Theory and computational details}

The details of the reference DFT calculations and the structure of the ML model closely follow those used in a recent paper focused on bulk structures\cite{lopa+23prm}, which allows us to perform an insightful comparative study. 
Here we provide a brief summary of these details and focus on the strategy we adopt to extend the training set to include disordered and inhomogeneous structures. 

\subsection{First-principles calculations}

The training data on energies and interatomic forces of the structures was obtained from density functional theory (DFT) simulations performed with the Vienna Ab initio Simulation Package (VASP)\cite{VASP}. 
The convergence criteria for the electronic self-consistent cycle was $10^{-6}$ eV, while the cutoff energy for the plane-wave basis set was 550 eV. 
We used a $\Gamma$-centered Monkhorst-Pack \cite{monk-pack76prb} scheme with a resolution of 0.04 $\pi$ \AA$^{-1}$ for sampling the first Brillouin zone. 
The behavior of the core electrons and their interaction with the valence electrons was described with projector-augmented wave (PAW) pseudopotentials \cite{kresse1999ultrasoft}. 
Exchange-correlation effects of the electrons were taken into account within the Generalized Gradient Approximation (PBEsol functional  \cite{csonka2009assessing}). 
The vacuum size for the surface slab calculations was set to 20 \AA. Following the same reasoning as in Ref. \citenum{lopa+23prm}, we performed the calculations without spin polarization.
\rev{Even though it is possible to describe the magnetism within the spin density functional theory (SDFT) approximation (and even though this framework is often used by ML models that incorporate magnetic information), when dealing with such a broad set of transition-metal compounds it is likely that some element combinations require different approaches, such as DFT with Hubbard U and J corrections (DFT+U+J), and dynamical mean-field theory (DMFT) \cite{Chen2016Spin, Huebsch2021Benchmark}. 
The use of different methods would introduce inconsistencies in the training data, and may also require the adjustment of external parameters such as Hubbard U and J for each individual structure. 
Hence, non-spin-polarized calculations remain the most consistent choice for this work, even though it limits the accuracy of the model for metals and alloys with strong magnetic behavior. 
We further discuss the errors associated with the neglect of magnetism in Sections \ref{sub:validation-surfaces} and \ref{sec:selected}, and in the Supplementary Information. }

\subsection{Machine-learning model}

For the sake of consistency between our previous work on bulk HEAs and our current work, we kept the model architecture unchanged. 
Here, we summarize the main ideas underlying the model, while for all the details, the reader is referred to the Ref. \citenum{lopa+23prm}.  

The model is based on a combination of ridge regressions built upon a chemical composition, two- and three-body correlation features, and a fully-connected neural network (NN) based on three-body features. 
We use the alchemical compression method introduced in Ref. \citenum{will+18pccp} to prevent the steep increase of model complexity due to the large size of the chemical space. 
The model connects a representation of the atomic environment $A_i$ of the atom $i$ with its contribution to the potential energy of the system $V(A_i)$:
\begin{equation}
    V(A_i) = V^{(aeb)}(A_i) + V^{(2B)}(A_i) + V^{(3B)}(A_i) + V^{(NN)}(A_i).
\end{equation}
Here, the first term $V^{(aeb)}(A_i) = \w^{(aeb)}_{a_i}$ correspond to an atomic baseline that depends only on the chemical identity $a_i$ of each atom.
The second term 
\begin{equation}
    V^{(2B)}(A_i) = \sum_{an}\w^{(2B)}_{a_ian}\rep <an||\field{\rho}_i^{1}>
\end{equation}
corresponds to a pair potential, that is trained in the full $25\times 25$-dimensional space of two-element correlations. 
The three-body potential
\begin{equation}
    V^{(3B)}(A_i) = \sum_{bnb'n'l}\w^{(3B)}_{bnb'n'l}\rep <bnb'n'l||\field{\tilde{\rho}}_i^{2}>,
\end{equation}
instead, is based on descriptors that are contracted from the original, 25-elements chemical space ($a$) to a much compressed pseudo-element space ($b$), 
\begin{equation}
\rep<b| = \sum_a u_{ba} \rep<a|.
\end{equation}
The coefficients $u_{ba}$ are collected in the alchemical coupling matrix $\mbf{u}$, that determine the compression from the full to the reduced chemical space. 
For consistency, we compress the chemical space to 4 pseudoelements, see Ref.~\citenum{lopa+23prm} for a discussion of convergence.  
The bra-ket notation represents in a concise manner the (al)chemical dimensions, as well as the indices of the radial basis ($n,n'$) and the angular momentum channel ($l$) that discretize the expansion of the neighbor-density correlations.  
A final NN term is designed to capture the non-linear, many-body components of the target potential:
\begin{equation}
    V^{(NN)}(A_i) = F\left[\{\rep <q||\field{\tilde{\rho}}_i^{2}>\}_q\right]. 
\end{equation}
For each environment, a multi-layer perceptron $F$ is applied to the set of compressed 3-body descriptors. 

\rev{
Although, as discussed above, we chose to build a non-magnetic model, one can see quite easily how this framework could be modified to incorporate some recent ideas to describe spin in a ML framework.
The simplest approach is to simply train the model on spin-polarized data with a fully relaxed spin subsystem. 
This approach provides an accurate description of a systems with a clear-cut magnetic ground state (such as the ferromagnetic \textit{bcc} iron \cite{Dragoni2018Achieving}), but is problematic when there are multiple near-degenerate spin states, e.g. near or above the Curie temperature.
Alternatively, one needs to modify the architecture to explicitly describe the magnetic state of the system, so that one can train on different magnetization states, and simulate atomic and spin dynamics simultaneously and thus provides a more accurate, and principled description of the system. 
This can be achieved by discretizing the spin states and treating different states as separate species\cite{eckh-behl21npjcm}, or by associating scalar descriptors with the magnetic state of the atoms \cite{Novikov2022Magnetic}. 
Both of these approaches are easily translated into the alchemically-compressed architecture.
A non-collinear, vectorial treatment of the atomic magnetic moments is also possible\cite{drau20prb,domi+22prb}, which would require more substantial modifications. 
As discussed above, in our opinion the challenge in including a consistent physical description of magnetism for such a wide chemical space is comparable -- if not harder -- than the implementation of a suitable ML architecture. 
}

\subsection{Training dataset construction}

The starting point for this study is the dataset of Ref.~\citenum{lopa+23prm}, which contains approximately 25000 bulk configurations, each containing between 3 and 25 transition metals (the entire $d$ block except for Tc, Re, Os, Cd, Hg) arranged as a regular, or randomly distorted, \emph{fcc} or \emph{bcc} lattice. 
We refer to this original dataset as ``subset O''. 
Although the resulting model was remarkably stable, and it was possible to perform molecular dynamics simulations of mixtures with 25 elements at temperatures well above the melting point, it would be unwise to use it for surface effects.
To obtain a model that is capable of a reliable description of the surface segregation process in HEAs with up to 25 \textit{d}-block elements, we need an extended data set, which we build iteratively by including three additional groups of structures.  

The first part of our extended dataset (labeled as ``A'') follows the same philosophy as the bulk data, but aims to capture surface energetics.
We generate a few subsets of on-lattice surface slab structures based on \textit{bcc} and \textit{fcc} lattices and Miller indices of  (100), (110) and (111). 
The unit cell of each structure contains 45 atoms, obtained by replicating the appropriately-oriented primitive cell of the surface according to a $3\times 3\times 5$ geometry. 
The atomic composition is randomly assigned by sampling from a random subset of elements ranging in size from 3 to 25. 
The lattice parameter is then rescaled according to the average atomic volume of the elements in the assigned composition.
For each subset of on-lattice data we generate a distorted counterpart by randomly displacing atomic positions and rescaling the lattice parameter. 
After generating 10000 samples  for each lattice and surface orientation, we select the 200 most diverse of them using a Farthest Point Sampling (FPS) algorithm \cite{imba+18jcp}, based on their two-body correlation features $\rep <an||\frho_i^{1}>$. 
Finally, for each combination of lattice and Miller index, we include 20 FPS-selected structures with $3\times 3\times 4$  and $3\times 3\times 6$ geometry, so that the ML model can capture finite size effects.
Overall, subset A contains contains 2640 surface slabs.  

In order to increase the structural diversity, we use replica-exchange molecular dynamics with atom swaps (REMD/MC) for structure generation and FPS for selection. These simulations use a preliminary version of the surface-enabled potential trained on a combination of subset A and the original dataset O. We perform high-temperature (2000 K - 4000 K) REMD/MC of 100 FPS-selected HEA bulk systems and intermediate-temperature (300K - 1500 K) REMD/MC of 100 FPS-selected HEA surface slabs. 
From the resulting trajectories, an additional round of FPS selection yields a collection of 1000 molten bulk structures (subset B) and 1000 thermally equilibrated surface slab structures (subset C).

Finally, using a preliminary potential trained on subsets O, A, B, and C, we generate a Cantor-style alloy surface slab containing all 25 elements in roughly equimolar proportions, perform intermediate-temperature REMD/MC sampling, and select by FPS a set of 500 structures (subset D) that mimics closely the setup we use to investigate the segregation propensity of different elements. All datasets used in this paper are summarized in Table \ref{tab:datasets}

\begin{table}
\centering
\begin{tabular}{  m{0.15\columnwidth}   m{0.6\columnwidth}   m{0.18\columnwidth}  } 
  \toprule[0.5pt]
  Label & Description & № Entries \\ 
  \midrule[0.5pt]
  O & Bulk structures & 25000\\ 
  A & Surface slabs & 2640 \\ 
  B & Bulk liquid, MD snapshots & 1000 \\ 
  C & Surface slabs, MD snapshots & 1000 \\ 
  D & Cantor alloy surface slabs, MD  & 500 \\ 
  \bottomrule[0.5pt]
\end{tabular}
\caption{Summary of the various training subsets we used to generate an extended HEA ML potential to study surface structures and energetics. Each subset is associated with a single-letter label which we use in the discussion.}
\label{tab:datasets}
\end{table}

\section{Transferability of alchemical learning}

The modeling scenario we investigate here -- the extension of an existing ML model to a substantially different type of configurations -- is common in practical applications to chemical and materials modeling.
The simplicity and interpretability of the architecture of the HEA25-4-NN model allows us to gain some generally applicable insights into successful strategies. 
For this reason, as well as because of the excellent performance of HEA25-4-NN in out-of-sample, extrapolative predictions, we chose to use exactly the same model architecture, despite the fact that recent results have shown that deep-learning models have the potential of improving substantially the in-distribution accuracy for the bulk HEA dataset\cite{pozdnyakov2023smooth}.

\begin{figure*}[tbp]
    \centering
    \includegraphics[width=0.9\textwidth]{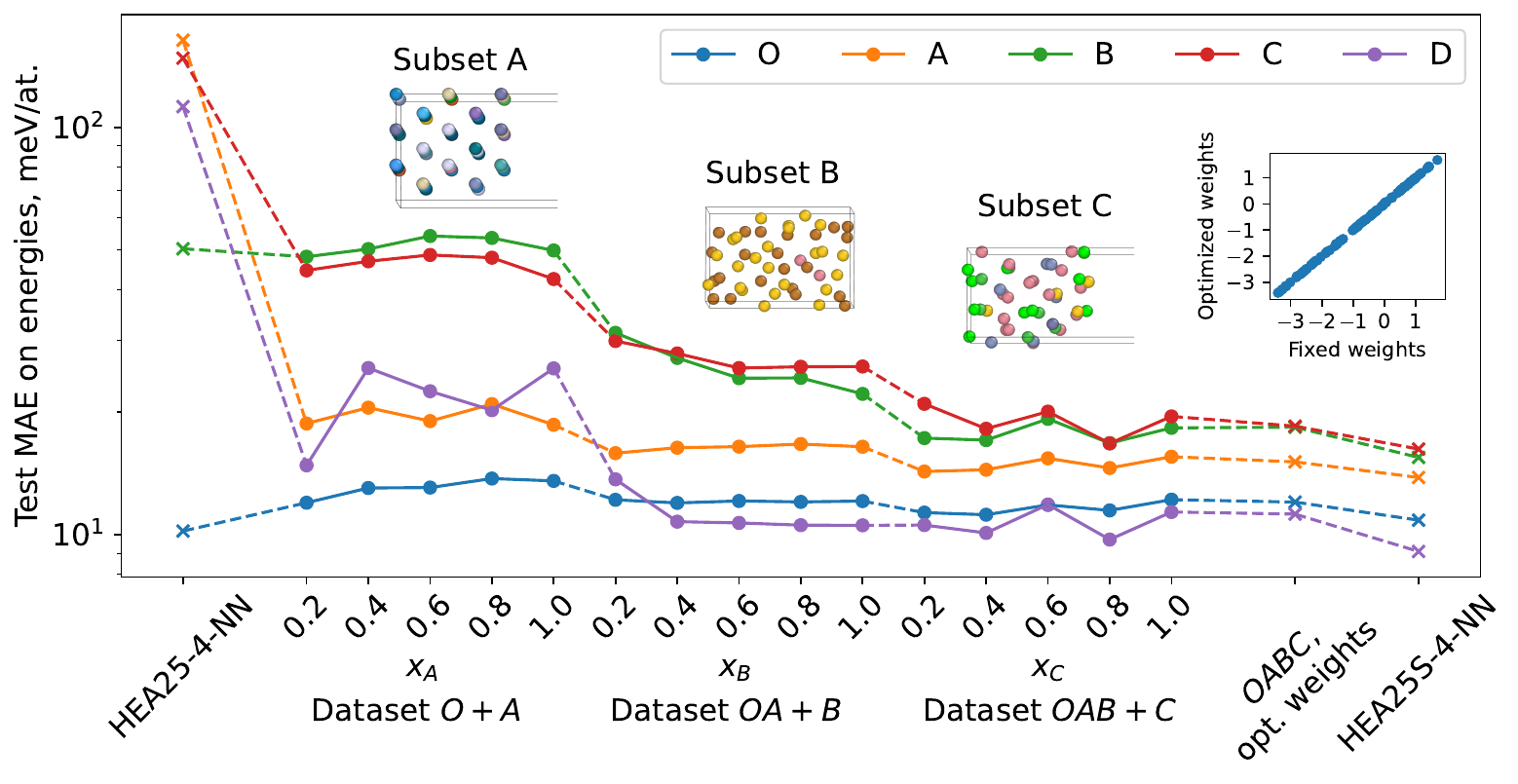}
    \caption{Learning curves for the validation error computed separately on the various datasets considered in this work. The errors are calculated on a hold-out test set containing 10\% of the structures. The labels for the datasets correspond to those in Table~\ref{tab:datasets}, and visualizations for each structure type are provided for convenience. For each curve, 10000 structures from the dataset O were included in the training set and kept fixed. 
    Each section on the \textit{x}-axis corresponds to the addition of a fraction of the structures for each subset ($x_A$ for the subset A, and so on), while the preceding datasets are included in full. 
    Individual data points marked with crosses correspond to the hold-out errors obtained by direct application of the HEA25-4-NN model from Ref. \citenum{lopa+23prm}, for a the model trained on the OABC dataset with optimized alchemical weights matrix, and the final HEA25S-4-NN model trained on the full dataset. The inset shows a parity plot of the alchemical coupling matrix $\mathbf{u}$ values in the case of model training with fixed $\mathbf{u}$  values (\textit{x}-axis) and optimizable $\mathbf{u}$  values (\textit{y}-axis).}
    \label{fig:lc_energies}
\end{figure*}

The extensive set of experiments in Ref.~\citenum{lopa+23prm} shows that the HEA25-4-NN model is able to perform both low-temperature and high-temperature MD without actually being trained on MD trajectories, and it even achieves respectable accuracy when making predictions for some elements that are not represented in the training set, thanks to the interpretable form of the alchemical contraction matrix.~\cite{lopa+23prm} 
Despite the overall stability, the accuracy for molten configurations or surface structures is much worse than for in-sample predictions, and so it makes sense to expand the training dataset to include more \emph{structural} diversity. 
An important question here, with a clear relevance for the broader goal of obtaining potentials that are generally applicable over a large swath of chemical space, is whether compositional \emph{and} structural degrees of freedom must be sampled independently.
In this particular situation, after having satisfactorily learned the behavior of 25 elements in the bulk phase, can we learn their behavior at a surface with a small number of additional training structures or do we need a similar amount of data for each type of chemical environment?

\subsection{Learning curves}

The predictions of the HEA25-4-NN model on the new structural subsets are up to 20 times less accurate than the in-sample predictions. While this might appear a very large degradation in performance, one has to consider that these structures are entirely different from the bulk structures included in dataset O, and indeed the errors on the high-temperature bulk structures are much smaller. 
In Figure~\ref{fig:lc_energies} we plot the accuracy of a series of models trained by progressively including data from each of the subsets A, B, C. When interpreting these curves, one should keep in mind that (1) the errors are computed separately on hold-out sets containing 10\% of the structures from all the available datasets, namely O, A, B, C, D;  (2) the alchemical weights matrix was taken from the HEA25-4-NN model and kept fixed during the training; (3) the models are trained incrementally, i.e. we start from the model weights of HEA25-4-NN and we further optimize the models after adding a new chunk of structures and after incorporating a new subset; (4) the train set is similarly extended in an incremental fashion when going left to right in the plot; (5) we always include 10000 randomly-selected structures from dataset O to avoid model drift. 
We also include the results of the HEA25-4-NN, the model trained on the full OABC dataset with optimized alchemical weights matrix (see the details in Sec. \ref{sec:aw_transferability}), together with results of the final HEA25S-4-NN model (see the details in the Sec. \ref{sec:final_model}) in Figure \ref{fig:lc_energies} to explicitly demonstrate the overall training dynamics. 

As the amount of training data increases, the accuracy of the model on this new data approaches the accuracy on the O subset. 
It is clear that the model can quickly adapt to new types of structures: adding the ``ideal'' surfaces from subset A improves the error by almost tenfold error not only on subset A, but also on the MD surface structures in subsets C and D; adding the bulk structures in subset B improves the accuracy on subset B, but also on the slab MD subsets C and D. 
The accuracy on dataset O degrades slightly initially, but the final model is only marginally worse than the one fitted exclusively on O-like structures. This indicates that, at least when including a substantial part of the original data in the new training, the model is flexible enough to incorporate new structures without substantially affecting the in-sample accuracy. 
We note however that learning curves exhibit a small slope which indicates that a much larger amounts of data would be needed to improve further the accuracy. 
Overall, this suggests a nuanced answer to our general question above. The trained model requires only a few structures to capture the gist of completely new types of configurations, and it can use the correlations from one set of structures to accelerate learning on new ones (for reference, the accuracy of the HEA25-4-NN model trained on 200 and 1000 O structures is approximately 150 and 30 meV/atom MAE\cite{lopa+23prm}). 
However, reaching uniform accuracy across completely different structural motifs may require almost-uniform train-set sampling. 

We also see that keeping fixed alchemical weights has a minimal impact on model accuracy, an observation we elaborate on in 
Sec. \ref{sec:aw_transferability}. %
From this we can conclude that all the general information about the interatomic interaction is already contained in the original dataset O, and one only needs to slightly adjust the model weights by adding a small amount of new data during training to reach semi-quantitative accuracy. 
\rev{The total number of structures in datasets A,B,C is less than 20\%{} of the 24'000 bulk structures that were used in Ref.~\citenum{lopa+23prm} to train the parent model -- and the learning curves indicate that an even smaller number may suffice with a limited impact on the accuracy.}
The relatively small change in model accuracy when adding the dataset D (which effectively corresponds to another iteration of a rudimentary active-learning protocol) is further indication that, within the level of accuracy we seek, the alchemical model can effectively combine information on different structure types, capturing 
the relevant information on surface effects from the dataset A and the defect properties of high-temperature molecular dynamics from the bulk dataset B to finally transfer this combined knowledge to the molecular dynamics of surface slabs. 

\subsection{Transferability of the alchemical weights matrix}\label{sec:aw_transferability}

The alchemical coupling matrix $\mathbf{u}$
is used to contract the power spectrum features from the original chemical space of 25 elements to a reduced-size alchemical space. 
In Ref.~\citenum{lopa+23prm} it was shown that the weights are relatively insensitive to the details of the fit, and that can be interpreted in terms of the layout of the elements in the periodic table. 
Here we can investigate further the generality of the contraction, by assessing 
how much it depends on the compositions of the training dataset. 
To study this, we start from the pre-trained HEA25-4-NN model and consider the dataset that combines the full O, A, B, C subsets, and perform two training exercises -  one with fixed alchemical weights and one in which the coupling coefficients are optimizable parameters.
The inset in Figure \ref{fig:lc_energies} shows that the entries of the coupling matrix remain almost unchanged upon optimization, which indicates that the values obtained from bulk structures are also compatible with the similarity in behavior of elements at surfaces, reinforcing the notion that  $\mathbf{u}$
contains valuable, interpretable and transferable information about the chemical relations of the elements across the periodic table. 
As one may expect, the small difference in alchemical weights results in negligible changes of the model predictions compared to the case of fixed alchemical weights. This is evident from the minute changes in test-set errors in Figure \ref{fig:lc_energies}, where the fixed-weights model is the last point in the ``OAB+C'' learning curve, and ``OABC, opt. weights'' indicates the model trained on the same dataset with optimizable alchemical coefficients. In fact, even the training curves of the two models are almost indistinguishable, as discussed in the Supplementary Materials.

\subsection{Final model and dataset}\label{sec:final_model}

Datasets O - C consistently sample the configuration space for simulations involving both bulk and surface-slab HEA configurations. 
Subset D serves mainly as a demonstration of convergence with respect to structural diversity. 
However, given that the application focus of this work is the study of surface propensity of different transition metals, we decided to train a final model that also includes these validation structures, allowing the weights of the alchemical coupling matrix to be optimized during the training to obtain the best possible accuracy with the current architecture. 
We refer to this combined dataset as the HEA25S dataset, and to this final model as the HEA25S-4-NN potential. Its accuracy on the hold-out test sets, computed separately for each subset, is presented as the last points of Figure \ref{fig:lc_energies}.
Inclusion of this last batch of structures leads to a slight improvement of the performance on all test subsets. The error of the model in predicting the energies of the HEA bulk structures, about 10 meV/atom, is effectively unchanged compared to our previous study \cite{lopa+23prm}. Errors on dataset D are comparable, around 9 meV/atom, and  those for subsets A-C are only slightly worse: from 13 to 16 meV/atom.  \todo{We make the final dataset and the model available as part of the data record that accompanies this publication\cite{hea25sdataset}.}

\begin{figure}[bthp]
    \centering
    \includegraphics[width=0.9\columnwidth]{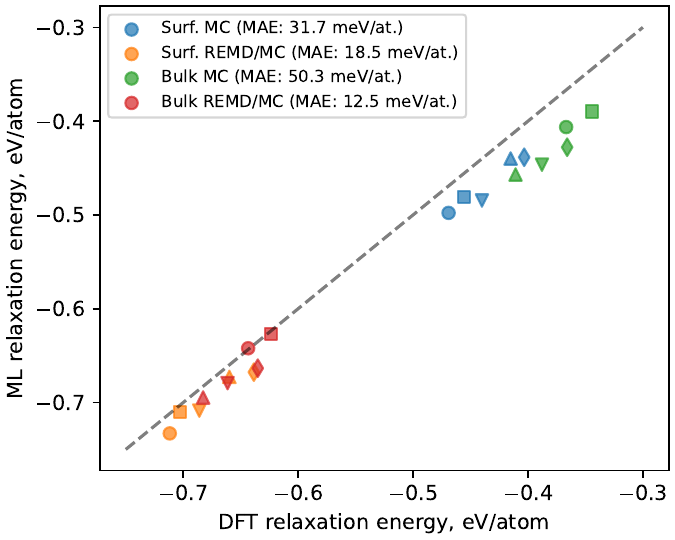}
    \includegraphics[width=0.9\columnwidth]{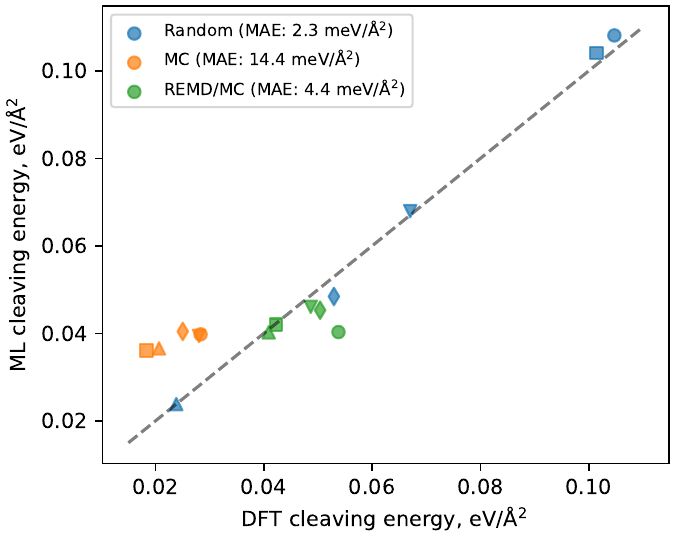}
    \caption{
\rev{    
Parity plots for relaxation (top) and cleaving (bottom) energies computed for 5 different realizations of a 25-element equimolar Cantor-style alloy structure. 
The symbols identify 5 different random realizations, and colors represent different simulation setups. MAE values in the legend provide an overall estimate of the error for each case. 
Relaxation energies are computed as the difference between the energy of an on-lattice, Monte-Carlo (MC) relaxed configuration, or fully-relaxed (REMD/MC) configurations of bulk and $(111)$ surface structures and that of a fully random, on-lattice configuration. 
Cleaving energies are computed as the difference between the energies of a bulk and a surface slab configuration, considering separately the cases of a fully-random, MC-relaxed or fully-relaxed configuration. 
}
}
    \label{fig:cantor-validation}
\end{figure}

\rev{
\subsection{Model validation on structure relaxation and surface processes}\label{sub:validation-surfaces}

In order to verify more quantitatively that the HEA25S-4-NN model can describe chemical and structural realaxation and surface effects, we generate a small set of more targeted test structures. 
We construct 
(a) five $2\times 5\times 10$ \emph{fcc} bulk structures of 100 atoms with 25-elements equimolar composition. 
Atoms are fixed to the ideal lattice positions, and the molar volume adjusted following the same logic used for the main dataset.  
Starting from these five structures, we use the HEA25S-4-NN model to generate: (b) bulk structures with the atomic ordering relaxed by Monte Carlo atom swaps at $T=300$~K, keeping a rigid lattice; (c) fully-relaxed bulk structures using combined REMD/MC sampling, followed by geometry optimization starting from uncorrelated low-temperature snapshots;
(d) $(111)$ slabs of the same size obtained by rigidly cleaving the random structures; (e) MC-relaxed $(111)$ slabs ; (f) fully-relaxed $(111)$ slabs.
For each configuration, we recompute the DFT energy, and the overall energy MAE for this set is  25.7 meV/atom -- with the largest errors being associated with on-lattice MC structures (see SI).

We then compute quantities that have a more direct bearing on relevant physical processes: Fig.~\ref{fig:cantor-validation}(top) shows a parity plot for ``relaxation energies'' computed as $E_b-E_a$, $E_c-E_a$, $E_e-E_d$, $E_f-E_d$, that quantify the enthalpic gain from short-range order, surface segregation and/or geometry relaxation. The subscript in these expressions corresponds to a specific subset from the list above. 
The ML model predicts these terms with an accuracy of 50.3, 12.5, 31.7, and 18.5 meV/atom respectively, which is less than 5\% of the typical relaxation energy on average.
We note that the typical error on MC relaxation energies prediction is significantly higher than in the case of REMD/MC relaxation. The main reason for such a difference is lack of the MC-relaxed structures in the training set. Therefore, the evaluation of the model on MC relaxation may be considered as another test for the model's generalization capabilities. 
Fig.~\ref{fig:cantor-validation}(bottom) shows a parity plot for the ``cleaving energies'' $E_d-E_a$, $E_e-E_b$, $E_f-E_c$. 
While these should not be interpreted to correspond exactly to a surface energy (averaging over multiple configurations, as well as incorporating entropic effects, would be necessary) they gauge the accuracy of the model for enthalpic terms that are directly associated with the creation of a surface. The accuracy (2.3, 14.4, 4.4 meV/\AA$^2$) is a fraction of less than 10\%{} the magnitude of the cleaving energy, except for the case of the on-lattice MC relaxed structures, that represent an extrapolative regime.
In the SI we also compare relaxation and cleaving energies computed with and without spin polarization. 
The errors are comparable to or smaller than the errors associated with the ML model, and small on the scale of the typical relaxation or cleaving energies, indicating that neglecting magnetic effects is an acceptable -- if harsh -- approximation when studying finite-temperature and surface effects on the structure and stability of HEAs. 
}

\section{Surface segregation propensity in a Cantor-style alloy}

\begin{figure}
    \centering
    \includegraphics[width=\columnwidth]{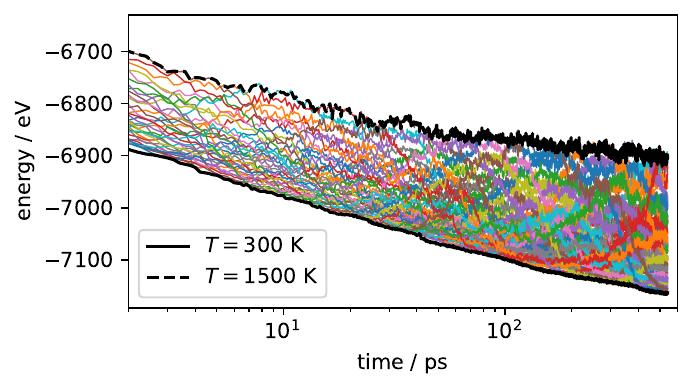}
    \caption{Typical trajectories of the potential energy for a 36-replica REMD/MC run of the 864-atom surface slab. Each colored line corresponds to one of the replicas, and the target temperature changes as Monte Carlo swaps occur. The full and dashed black lines are obtained by concatenating the segments that correspond to the lowest and highest simulation temperature. Note the logarithmic $x$ axis. }
    \label{fig:cantor_remd_traj}
\end{figure}

In heterogeneous catalysis, the composition and structure of the surface are among the most important factors determining the activity of a material. 
In this respect, HEAs are interesting because they often exhibit differential segregation of their components at the surface\cite{Ferrari2020, Wynblatt2019, Dahale2022, Chatain2021, Middleburgh2014}, which can be used to control their properties. 
For example, this effect can be exploited to use small quantities of a rare-earth element, and manipulate the alloy chemistry so that it accumulates at the surface, where it can enhance the catalytical activity.\cite{Katiyar2021, Guisbiers2016, Xin2020, Pedersen2020, Pedersen2021-2}.
Even though the actual segregation propensity will depend on the specific composition, as well as on processing conditions that might affect thermodynamics and kinetics, it is useful to establish an ``absolute'' scale for the surface affinity of each of the 25 transition metals we consider in this study. 
To do so, we perform simulations similar to those performed in Ref.~\citenum{lopa+23prm} for a bulk system. 
We consider a surface slab of a Cantor-style alloy, containing all 25 elements in equimolar proportions, randomly assigned to the atomic sites of an 864-atom slab with an initial \textit{fcc} lattice, oriented to expose a (100) surface.
We then perform four independent REMD/MC runs, combining replica-exchange molecular dynamics (we use 36 replicas with temperatures distributed from 300 K to 1500 K on a logarithmic scale) with Monte Carlo swaps between atom types. 
Molecular-dynamics details are the same as in Ref.~\citenum{lopa+23prm} with a time step of 2 fs and an efficient Langevin thermostat. The cell is left free to fluctuate along the in-plane directions, with a small applied pressure of 1 bar. The system shows glassy behavior, with rapid disruption of the \textit{fcc} lattice, and logarithmic relaxation of the potential energy (Fig.~\ref{fig:cantor_remd_traj}).
\rev{This loss of crystallinity is not due to the presence of a surface, but a consequence of the presence of a wide variety of species with very different atomic radii, and was also observed in bulk simulations\cite{lopa+23prm}.}
The simulation is not fully equilibrated even after 500~ps and several hundred thousands attempted atom swaps. We observe however that the four trajectories behave in a similar way, and that structure and composition profiles remain roughly constant after a few hundred ps.  %
Thus, we average over the last 100 ps of each of the four independent runs to provide a qualitative estimate of the surface affinity.

\begin{figure}
    \centering
    \includegraphics[width=\columnwidth]{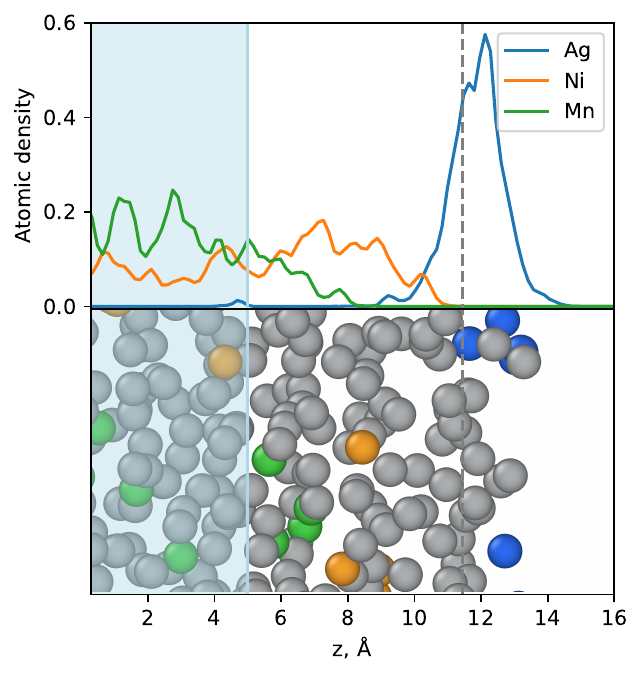}
    \caption{Concentration profiles of a few representative elements (Ag, Ni, Mn) averaged on the last 100 ps of four REMD/MC simulations of the Cantor-style alloy at 300 K with different surface segregation patterns (top) and a single snapshot of the system taken from the end of the relaxation trajectory (bottom). The z-coordinate represents the distance from the center of the surface slab, while the light blue area on the figure and a dashed line represent the bulk region and the Gibbs surface plane respectively. Ag, Ni, Mn atoms are color-coded consistently with the concentration profiles in the upper panel, while all other atoms are shaded in gray.}
    \label{fig:cantor_profile}
\end{figure}

We compute and analyze the concentration profiles of different elements along the normal direction to the surface plane. 
Fig. \ref{fig:cantor_profile} showcases a few representative cases, together with a snapshot of a longitudinal cut of the simulation slab.
Concentration profiles for all elements and different temperatures are given in the SI.
There are clear-cut differences in the surface segregation propensities, with elements such as Ag accumulating at the surface, others such as Mn being strongly depleted, and some such as Ni displaying an intermediate behavior.  The concentration profiles show pronounced differences in the peak shapes and positions, which suggests a tendency for some elements to accumulate in the sub-surface layer and to form ordered surface alloys. 
However, the disordered structure, the slow relaxation and the limited accuracy of the potential make it difficult to provide definitive statements on the precise relaxed geometry of the structure that is formed at the surface.
Instead, we compute an integrated, macroscopic indicator of segregation propensity in terms of the Gibbs surface excess $\Gamma_a$, that is defined for each element $a$ as
\begin{equation}
    \Gamma_a = \frac{N_a - N_a^{B} \cdot N / N^B}{S}, \label{eq:gibbs-excess}
\end{equation}
where $N_a$ and $N_a^B$ correspond to the total number of atoms in the slab, and the number of atoms inside the bulk region of the surface slab. $N$ and $N^B$ represent the total numbers of sites in the simulation and inside the bulk region, respectively, and $S$ is the surface area. We define the bulk region of a surface slab as a 10\AA{}-thick region around the center of the slab (shown as a shaded region in Fig.~\ref{fig:cantor_profile}). 
Values of $\Gamma_a > 0$  indicate the tendency of a given element to accumulate at the surface. If $\Gamma_a \simeq 0$ , then the element does not have a pronounced affinity, and will have roughly the same concentration in the bulk and at the surface. Finally, if $\Gamma_a < 0$, the element will tend to stay in the bulk region, and be depleted at the surface.
It is important to stress that the definition~\eqref{eq:gibbs-excess} is insensitive to the thickness of the bulk region, which is only used to estimate the concentration of each element ``far'' from the surface. It avoids specifying explicitly which layers are considered as part of the surface, and it provides an overall measure of the surface affinity that can be readily connected to macroscopic quantities, such as the surface energy\cite{adam41book}. 

\begin{figure}
    \centering
    \includegraphics[width=\columnwidth]{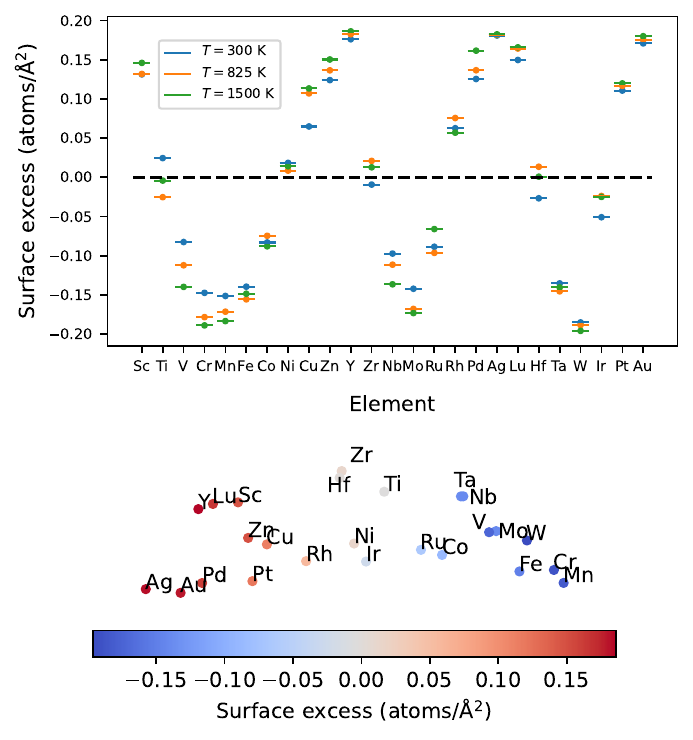}
    \caption{Values of the surface excess $\Gamma_a$ at different temperatures for each of 25 elements from the Cantor-style alloy REMD/MC simulation (top). The element similarity map from our previous work \cite{lopa+23prm}, based on simulation of bulk Cantor-style alloy of the same composition at 1253 K and color-coded according to the surface excess values at the same temperature from this work (bottom). }\label{fig:surface_excess_cantor}
\end{figure}

Figure~\ref{fig:surface_excess_cantor}a shows the computed surface excesses for the 25 elements at three representative temperatures. Differences are very pronounced, with elements at the edges of the \textit{d} block showing surface affinity, and most of the elements in the central part of the $d$ block being depleted at the surface. 
The temperature dependency of the surface excess is usually weak, except for a few cases such as V or Nb, where it changes by more than 50\%{}  between 300~K and 1500~K. 
There is a remarkable correlation between the surface excess of the different elements and their position on the 2D map constructed in Ref.~\citenum{lopa+23prm} based on their bulk behavior (Figure~\ref{fig:surface_excess_cantor}b). 
This is perhaps unsurprising, given that the mutual affinity in the bulk and the affinity for the surface are both loosely related to the binding energy between atoms.
\rev{
It is also worth noting that the periodic trends observed for the surface excess, as well as the ordering derived from bulk short-range-order calculations, reflect the trends in atomic radii across the transition metals, with larger radii at the beginning and the end of each period.
This is consistent with the fact that transition metals surfaces are characterized by an accumulation of tensile stress\cite{Shiihara2013Origin,Shiihara2016Contribution}, that is alleviated by the segregation of elements with a large radius. 
}
Overall, this observation reinforces the notion that \rev{the data-driven optimization of alchemical weights can (re)discover physical trends across the periodic table,\cite{will+18pccp} and that similarity in behavior in REMD/MC simulations driven by the HEA25S-4-NN model } can be used as a guide to performing element substitutions when designing novel alloy compositions. 

\section{Surface segregation in selected alloys}\label{sec:selected}

The surface segregation study in the previous section is intended to provide insight into the overall surface propensity of transition metals commonly found in HEAs.
However, it cannot substitute an explicit study of specific compositions, nor does it provide validation for our computational framework.  
Ideally, one could assess the reliability of our framework  by  directly comparing the equilibrium surface composition for specific HEA compositions with the surface distribution of elements observed in experiments.
Nevertheless, there are some key concepts that need to be considered when evaluating our results. 
First, the model is trained on DFT data and thus the accuracy is limited by the accuracy of the DFT reference \rev{-- which notably, for our model, neglects effects due to magnetism.} 
Second, our sampling protocol allows to achieve highly ergodic sampling, whereas kinetic trapping plays a key role in (meta)stabilizing HEAs. 
Third, we consider the HEA surface slabs under vacuum conditions, at variance with experimental setups where a significant amount of oxygen is often present. 
In fact, the presence of a chemically-active environment can dramatically affect the surface concentration of different species,  due to the formation of oxides on the surface (see Ref. \citenum{Ferrari2020}) or the leaching of some of the elements in solution.
In this section, we first validate the ML potential against DFT calculations for the prototypical Cantor alloy CoCrFeMnNi and comment on the suitability of ``static'' approaches to evaluate segregation patterns, which are commonly  used in combination with first-principles calculations \cite{Ferrari2020}.   
Then, we perform simulations for an IrFeCoNiCu alloy and comment on the comparison with experimental data on the surface composition from Ref. \citenum{Maulana2023}. 

\subsection{CoCrFeMnNi alloy}

We begin by ensuring that the accuracy of the HEA25S-4-NN model is sufficient to reproduce the energetics of surface segregation relative to reference DFT calculations. 
To do so, we apply a simplified version of the protocol used in Ref.~\citenum{Ferrari2020}: we generate multiple slabs using the ideal lattice, an equimolar composition and random element distribution, and compute the difference in enthalpy when all the surface sites are filled with a specific atom type X=\ce{Cr,Mn,Fe,Co,Ni}, keeping the overall composition fixed: $H_\text{segr} = H_{\text{X,surf}} - H_\text{rand}$. This quantity is averaged over 16 different random realizations.  

\begin{figure}
    \centering
    \includegraphics[width=\columnwidth]{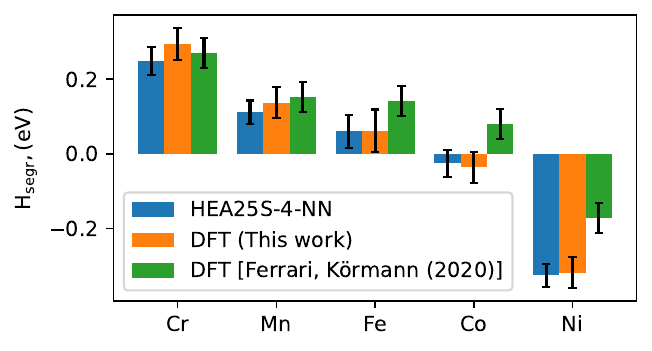}
    \caption{Segregation enthalpies per surface atom of CoCrFeMnNi elements at 300 K, calculated with the  HEA25S-4-NN potential (blue) and with the reference DFT setup (orange). Non-magnetic PAW DFT data on Gibbs segregation energy from Ref. \citenum{Ferrari2020} is provided for comparison (green). The error bars for HEA25S-4-NN and DFT results represent the standard deviation of the values obtained from 16 independent calculations. The reference DFT data uncertainty of 40 meV is taken from Ref. \citenum{Ferrari2020}. }
    \label{fig:CoCrFeMnNi_G_segr}
\end{figure}

As shown in Fig.~\ref{fig:CoCrFeMnNi_G_segr}, there is excellent agreement between the values of segregation enthalpy obtained using the ML potential and those obtained with explicit DFT calculations. 
Results also agree qualitatively with Ref~\citenum{Ferrari2020}, with Ni being the element with the highest surface propensity (large negative $H_\text{segr}$) and Cr that with the least propensity (large positive $H_\text{segr}$). Mn and Fe have a small positive $H_\text{segr}$, while Co has a negative segregation enthalpy (i.e. slightly favors surface sites), at variance with Ref.~\citenum{Ferrari2020}, in which it was found to have a (very small) positive $G_\text{segr}$.   
The quantitative discrepancy can be attributed to the difference in DFT details and to the fact that Ref.~\citenum{Ferrari2020} includes entropic terms and computes the segregation energy at constant bulk composition, which requires estimating the chemical potential of the elements. 
We chose to ignore these terms to obtain a more transparent validation of the ML model. We discuss further the details of the calculation of $H_\text{segr}$ in the Supplementary Information.

Having access to a fully-flexible and inexpensive ML model, we can then verify whether estimates based on static calculations, and on the overly simplified picture of the formation of a pure surface monolayer on top of a random alloy, provide reliable indications of actual surface propensity. 
To do this, we prepare a surface slab with a \textit{fcc} lattice, (111) orientation, and a  larger $7\times 7\times 11$ supercell containing 539 atoms.
We perform a REMD/MC run of this surface slab for 200 ps in the NPT ensemble using a 2 fs  timestep, following the same workflow we applied to the 25-elements Cantor alloy. 
As shown in the SI, the resulting REMD/MC trajectories reach equilibrium very quickly, which allows performing a single run, analyzing the surface composition by averaging over the final 100 ps of the trajectory. 
Even the replicas at the highest temperature retain an ordered \emph{fcc} structure, consistent with the fact that the simulations are performed below the melting point. 
\rev{
To complement this analysis, we show in the SI the values of $H_\text{segr}$ computed with spin-polarized DFT, and compare them with the mean enthalpy of structures that are relaxed with MC sampling on a rigid lattice at 300~K using the (non-polarized) HEA25S-4-NN model, and with combined REMD/MC sampling, also at 300~K. 
Neglecting finite-temperature sampling of short-range order and structural relaxation is an approximation that is at least as severe as neglecting effects due to magnetism. 
Compositional and structural relaxation stabilize greatly (up to several hundred meV/atom) the slab, in comparison to the static calculations in Ref.~\citenum{Ferrari2020}.
This is also true when considering spin-polarized DFT calculations, even though the relaxation procedure neglects magnetic effects.
In other terms, performing a full relaxation using a non-magnetic ML model leads to a better prediction of surface segregation incorporating magnetism \emph{a posteriori} than using spin-polarized DFT calculations limited to highly-idealized static structures. 
ML potentials that are capable of describing atomic spins\cite{eckh-behl21npjcm,Novikov2022Magnetic,domi+22prb}, in combination with thorough thermodynamic sampling, would obviously provide an even better approximation -- even though extending spin-aware models to a similarly broad chemical space raises conceptual issues on the most appropriate physical description of magnetism, and even though finite-temperature effects are likely to reduce the importance of spin correlations for HEAs, that usually have low Curie temperatures \am{  \cite{Wang2021Nanocrystalline}. }
}

\begin{figure}
    \centering
        \includegraphics[width=\columnwidth]{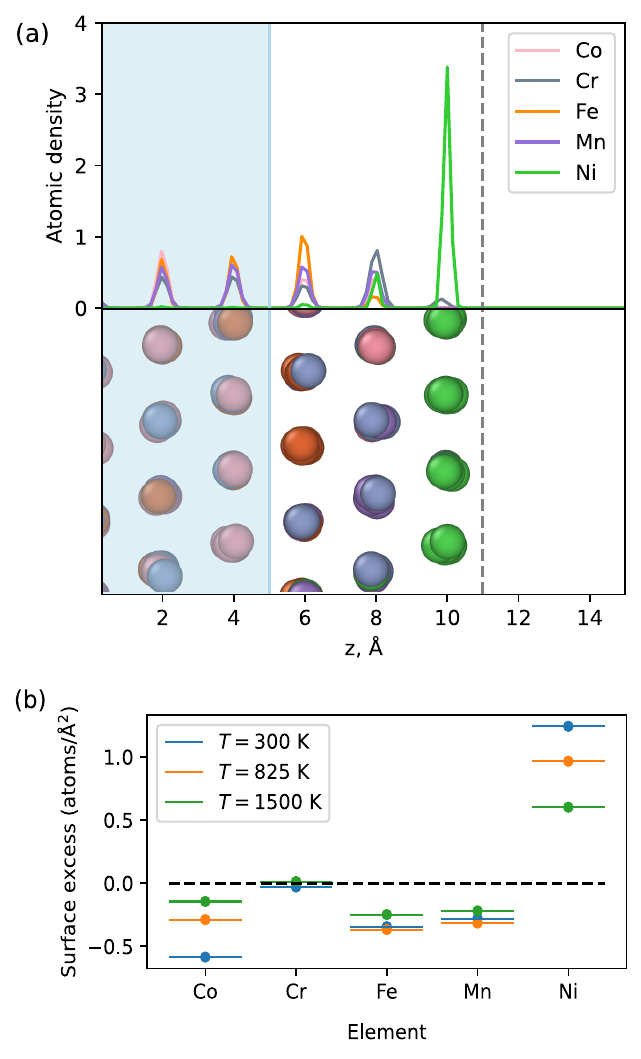}
    \caption{(a) Concentration profiles of the elements from the REMD/MC simulation of the CoCrFeMnNi alloy at 300 K (top) and a single snapshot of the system taken from the end of the relaxation trajectory (bottom). The z-coordinate represents the distance from the center of the surface slab, while the light blue area on the figure and a dashed line represent the bulk region and the Gibbs surface plane respectively. Atoms in the snapshot and concentration profiles follow the same color scheme.
    (b) Gibbs surface excess $\Gamma_a$ at different temperatures for the elements in the CoCrFeMnNi alloy from the REMD/MC simulations.}
    \label{fig:CoCrFeMnNi_profile_and_surface_excess}
\end{figure}

Analyzing more carefully the surface structure for ML-driven REMD/MC calculations, we observe that the concentration profiles of the elements at 300 K (see Figure \ref{fig:CoCrFeMnNi_profile_and_surface_excess}a) show a very strong nickel enrichment in the first surface layer and depletion for all other elements. 
The second layer contains a significant excess of chromium, while iron, manganese and cobalt are more abundant from the third layer and deeper into the bulk.
These observation are reflected in the Gibbs surface excess, computed  according to Eq. \eqref{eq:gibbs-excess} (see Figure \ref{fig:CoCrFeMnNi_profile_and_surface_excess}b).
There is a large surface excess for Ni and strong depletion of Co, which becomes less pronounced at the highest temperature. 
Mn and Fe have weak depletion, and Cr has almost zero overall surface activity -- even though, as evidenced by Fig.~\ref{fig:CoCrFeMnNi_profile_and_surface_excess}b, this is the net result of first-layer depletion and second-layer accumulation. 
When comparing these observations with the statically-determined $H_\text{segr}$, one sees that the strong surface affinity of Ni is consistent between the two approaches \rev{and between spin-polarized and non-polarized calculations,} as is the fact that Mn and Fe have better affinity for the bulk. 
The large positive $H_\text{segr}$ of Cr does not translate in strong depletion, which we attribute both to the fact that Cr appears to find a favorable environment in the subsurface layer, and to the fact that Cr exhibits strong local ordering in the bulk (as observed in Ref.~\citenum{lopa+23prm}), which makes a completely random alloy a poor reference state for the calculation of $H_\text{segr}$. 
Similarly, our static DFT calculations show negative $H_\text{segr}$  for Co (indicating surface propensity) but finite-temperature sampling shows a negative surface excess $\Gamma_\text{Co}$ (indicating surface depletion). 
These results demonstrate the necessity of performing simulations that allow sampling local bulk relaxation and the formation of non-trivial surface orderings, as well as how drastically the behavior of the elements may differ from static, highly-idealized reference calculation.

\subsection{IrFeCoNiCu alloy}

\begin{figure}
    \centering
        \includegraphics[width=\columnwidth]{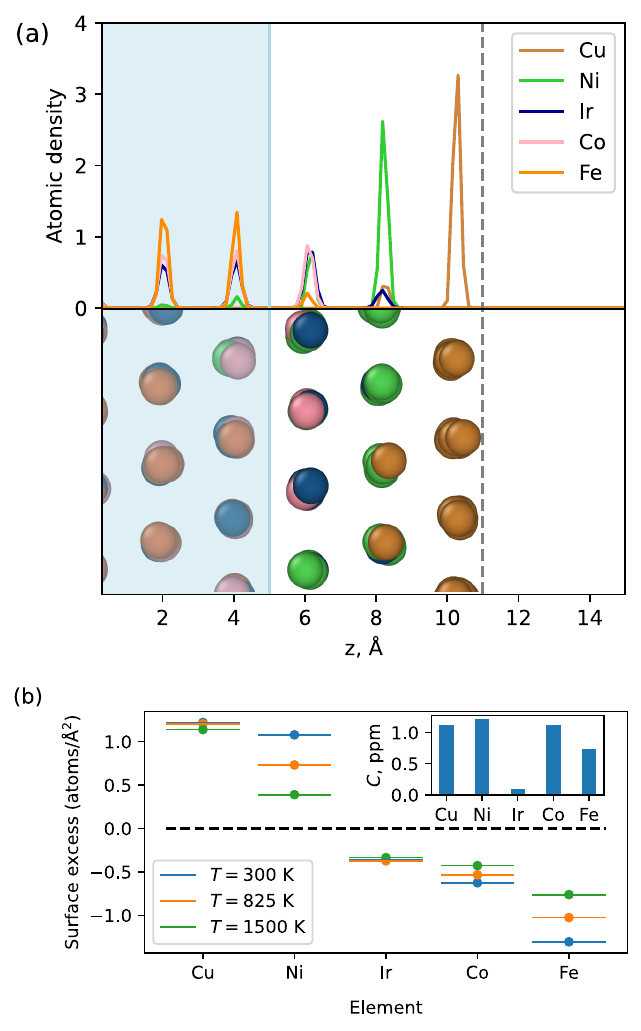}
    \caption{(a) Concentration profiles of the elements from the REMD/MC simulation of the IrFeCoNiCu alloy at 300 K (top) and a single snapshot of the system taken from the end of the relaxation trajectory (bottom).  The z-coordinate represents the distance from the center of the surface slab, while the light blue area on the figure and a dashed line represent the bulk region and the Gibbs surface plane respectively. Atoms in the snapshot and concentration profiles follow the same color scheme.
    (b) Gibbs surface excess $\Gamma_a$ at different temperatures for the elements in the IrFeCoNiCu alloy from the REMD/MC simulations. Experimental data on the concentration of dissolved metals $C$ after electrochemical treatment of the IrFeCoNiCu nanoparticles in a 0.1 M HClO$_4$ solution under the applied voltage from Ref. \citenum{Maulana2023} is provided in the inset for comparison.}
    \label{fig:IrFeCoNiCu_profile_and_surface_excess}
\end{figure}

Even though obtaining a realistic description of local ordering is necessary to extract reliable estimates of thermodynamical properties, experimental conditions often require a description of kinetic effects and of the presence of a reactive environment at the surface. 
To investigate these effects, we consider the case of an equimolar IrFeCoNiCu HEA, for which a detailed experimental characterization of the surface composition has been recently described.\cite{Maulana2023}
We first compute the Gibbs surface excess following the same protocol we used for the CoCrFeMnNi alloy. In this case, simulations show high segregation propensity of Cu and Ni (Fig. \ref{fig:IrFeCoNiCu_profile_and_surface_excess}). Cu atoms are predominantly found in the first surface layer, while Ni atoms are mostly found in the second layer (Fig. \ref{fig:IrFeCoNiCu_profile_and_surface_excess}). The remaining elements have negative surface excess and segregate to the bulk, with iridium having the smallest absolute value of $\Gamma$.
We also observe that increasing temperature does not change the relative order of segregation propensity of the elements, but is associated with a more even distribution of the elements across the cell at high temperatures. 
\rev{Similar to the case of CoCrFeMnNi, spin-polarized DFT calculations confirm the qualitative findings, with both on-lattice and fully-relaxed structure being much more stable than the random reference even when re-computed including magnetic effects. 
Even though the quantitative value of the segregation enthalpy is considerably smaller, there is no indication of a major discrepancy that might alter fundamentally the trends we observe with the non-magnetic ML potential.
}

\begin{figure}
    \centering
         \includegraphics[width=\columnwidth]{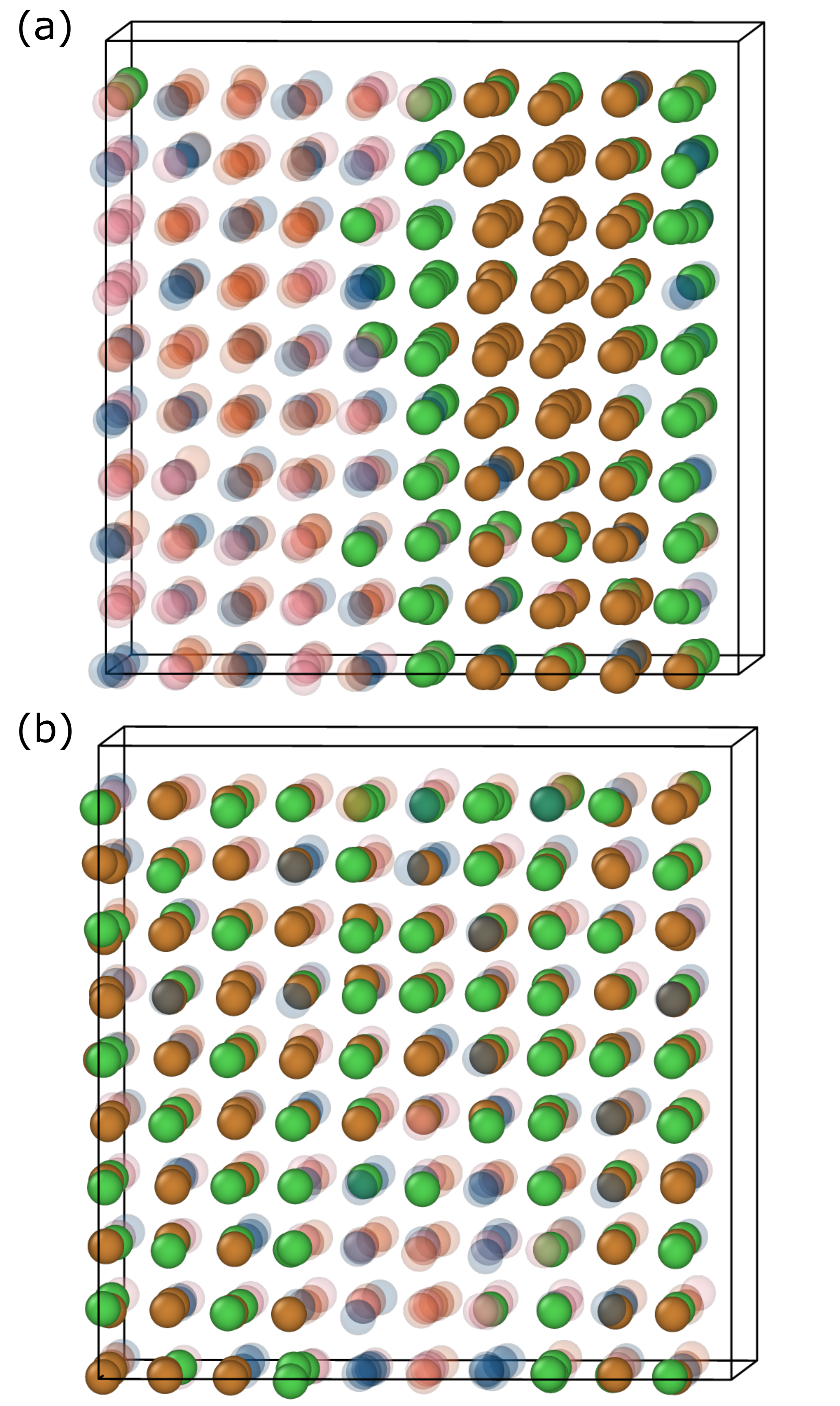}
    \caption{Front (a) and side (b) views of the IrFeCoNiCu bulk system, relaxed at 300 K in the REMD/MC simulation. Brown and green spheres represent Cu and Ni atoms, while the remaining elements are transparent. Precipitation of Cu and Ni is observed after relaxation, which indicates instability of the alloy.}
    \label{fig:IrFeCoNiCu_bulk}
\end{figure}

The comparison of these results with the experimental data in Ref. \citenum{Maulana2023}, shown in the inset,  requires a brief description of the experimental setup. First, the IrFeCoNiCu nanoparticles synthesized in that work were treated with a 0.1 M solution of perchloric acid (HClO$_4$) at room temperature under an applied voltage, which essentially washed away the outer shell of the nanoparticle. 
Scanning transmission electron microscopy with energy dispersive spectroscopy imaging reveals an iridium-rich region at the surface. 
As evidenced by the high concentration of all elements but iridium found in the solution, this process is driven by the reactivity of the metals in the aggressive chemical environment, rather than by an intrinsic tendency of Ir to accumulate at the surface. 

\begin{figure}
    \centering
        \includegraphics[width=\columnwidth]{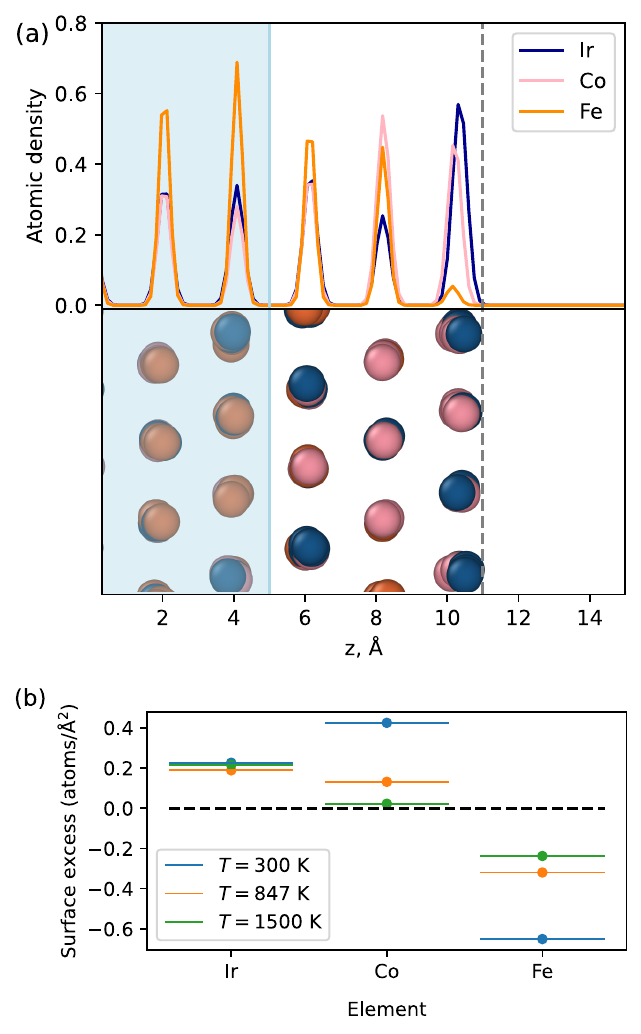}
        
    \caption{(a) Concentration profiles of the elements from the REMD/MC simulation of the IrFeCo alloy at 300 K (top) and a single snapshot of the system taken from the end of the relaxation trajectory (bottom). The z-coordinate represents the distance from the center of the surface slab, while the light blue area in the figure and a dashed line represent the bulk region and the Gibbs surface place respectively. Atoms in the snapshot and concentration profiles follow the same color scheme. 
    (b) Gibbs surface excess $\Gamma_a$ at different temperatures for the elements in IrFeCo alloy from the REMD/MC simulations. }
     \label{fig:IrFeCo_profile_and_surface_excess}
\end{figure}

Our simulations suggest that in the absence of an electrochemical treatment, thermodynamic drive would favor the formation of a Cu/Ni thin layer at the surface. 
This suggests a broader investigation of the stability of the IrFeCoNiCu alloy, which we tackled by performing a  REMD/MC simulation of a 5x5x5 bulk \textit{fcc} crystal supercell (500 atoms in total), over the same temperature range (300 K - 1500 K) used for the slab, and at a constant pressure of 1 bar.  
Our results (Fig. \ref{fig:IrFeCoNiCu_bulk}) suggest that the alloy is thermodynamically unstable, leading to precipitation of Cu and Ni even in the bulk phase. 
Therefore, we expect the homogeneous IrFeCoNiCu nanoparticles from Ref. \citenum{Maulana2023} to be only metastable. The sluggish diffusion that kinetically stabilizes the quasi-random alloy may also affect the surface stability, explaining why the surface remains relatively homogeneous in the absence of a chemical treatment. 
However, the instability of the bulk alloy and thermodynamic drive of Cu and Ni to segregate at the surface may become a serious problem in catalysis applications, by reducing the surface concentration of Ir over time, especially if operating temperatures are reasonably high and if the environment does not ensure continued leaching of the reactive metals.

It would be desirable to have an alloy that is both thermodynamically stable in bulk form and that shows an intrinsic drive towards surface accumulation of Ir. 
While chemical treatment might still be useful to accelerate the formation of the Ir layer, such alloy would likely be easier to manufacture and more stable in practical applications. 
Based on the observation that the Ir-Fe-Co phase remains stable in our REMD/MC simulation and the relatively higher surface propensity of Ir over Fe and Co, we investigated the ternary IrFeCo system, which was recently patented as a promising candidate for applications in catalysis \cite{IrFeCo_patent}. 
We further verify the bulk stability of the IrFeCo by performing exactly the same REMD/MC simulation as for the 5-element composition. In this case, we observe that the relaxed structure remains uniformly mixed even after convergence of the trajectories (see the SI). 
Given that the highest temperature in the REMD/MC run is 1500 K, we expect the alloy to preserve its phase stability also at high temperature. 
Furthermore, we also study surface segregation with a setup analogous to that used for the other systems: we create a 7x7x11 \textit{fcc} surface slab and perform a REMD/MC simulation in this system over the same range of temperatures from 300 K to 1500 K and at constant pressure of 1 bar. 
We then calculate the surface excess using Eq.~\eqref{eq:gibbs-excess} after averaging the composition of the bulk region over last 100 ps of the trajectory. The results are presented in Figure \ref{fig:IrFeCo_profile_and_surface_excess}.   
Ir tends to accumulate at the first surface layer, while Co is abundant in both the first and second layer. 
As a result of the balance between the propensity of the two elements for the surface and sub-surface layer, the overall value of $\Gamma_\text{Co}$ shows a very pronounced temperature dependence. 
At 300 K Co is the most surface-active, followed closely by Ir, while at higher temperature Ir remains abundant in the first layer, while the concentration of Co becomes more uniform, reducing its overall excess to almost zero. 
The bulk phase stability and the positive $\Gamma_\text{Ir}$ at all temperatures make the IrFeCo system potentially interesting for applications, allowing one to create nanoparticle with a thermodynamically-stable Ir surface layer.

\section{Conclusions}

We extend a recently-developed machine-learning potential trained on the DFT energetics of arbitrarily complicated alloys of 25 \emph{d}-block metals to include surfaces and defective structures, allowing us to study equilibrium surface segregation of the various components. 
By investigating in detail the effect of incorporating new structures and re-optimizing a pre-trained model, we infer some general guidelines for the broadly-relevant goal of training models for chemically and structurally diverse problems. 
First, it appears that the ``alchemical embeddings'' we use to reduce the complexity of learning across a large sector of the periodic table are transferable between different types of structural environments. 
The weights that describe how individual elements are projected on a smaller-dimensional set of pseudo-elements change only minimally if we re-optimize them on the extended data set, and the impact on the model accuracy is negligible. 
We also observe that very few structures need to be added to improve the accuracy for a new class of configurations: whereas the bulk-only model reached saturation with approximately 10000 structures, a few 100s of surface and molten configurations are sufficient to improve the validation error by almost an order of magnitude. 
On the other hand, the error saturates at a semi-quantitative level of accuracy, with errors between 10-20 meV/atom that are too large to resolve the fine details of the energetics of intermetallic phases. 
It appears that more flexible models would be necessary to reach quantitative levels of accuracy\cite{pozdnyakov2023smooth}, although this may come at the price of reduced extrapolative power, and an increase in the data requirements. 

The new model, which we name HEA25S-4-NN, is then used to thoroughly sample configurations and element distributions for a large slab of a 25-elements Cantor-style alloy. Based on the quasi-stationary state of these simulations (that never fully achieve equilibrium due to the glassy nature of the system) we compute the Gibbs surface excess of all 25 elements. 
This quantity, that can be interpreted as a measure of their surface propensity, shows clear periodic trends, \rev{that are consistent with the known surface stress effects that would favor the presence at the surface of atoms with a large metallic radius,} and correlates very strongly with a measure of chemical similarity that was derived in Ref.~\citenum{lopa+23prm} based exclusively on bulk short range ordering information.  
Thus, one can use the bulk-derived similarities as a guide for alloy design also in terms of the surface activity of the various components.

We then present two examples of the application of the HEA25S-4-NN model to the study of surface segregation of equimolar quinary alloys. 
We perform a direct validation of the accuracy of the model against a DFT assessment of the surface segregation enthalpy of different species in CoCrFeMnNi, finding errors that are well below 10\%{} of $H_\text{segr}$. 
Furthermore, we demonstrate that the kind of static, idealized calculations that are usually employed to investigate surface propensity with explicit electronic-structure calculations can lead to qualitative errors, such as estimating a strong tendency towards surface depletion of Cr, whereas explicit REMD/MC sampling shows a tendency for Cr to accumulate in the sub-surface layer.
\rev{ The errors associated with treating the HEAs as random and/or rigid lattices are comparable (and usually larger) than those associated with the neglect of magnetism, which is one of the most prominent physical limitations of our ML model. }
At the same time, one has to acknowledge that simulations that aim to compute the equilibrium thermodynamics of a solid-vacuum interface cannot directly describe more complicated experimental or \emph{in operando} conditions. 
Our simulations of a IrFeCoNiCu suggest that the bulk alloy is only metastable, and tends to phase separate into a NiCu and an IrFeCo phases. The instability is also reflected in the strong tendency of Cu and Ni to form a surface bilayer. 
The accumulation of Ir at the surface of seemingly stable quinary alloy nanoparticles that has been recently observed in experiments\cite{Maulana2023} can only be understood in terms of the slow diffusion kinetics in the bulk phase, and the leakage of chemically active elements in the aggressive environment used to treat the particles. 
Based on our simulations, we propose that the ternary IrFeCo alloy may be a more stable composition, that also exhibits a thermodynamic drive to form a stable Ir surface layer at all temperatures.

This study demonstrates the ease by which ML potentials for chemically-diverse datasets can be extended to include new types of structures, and used to capture the qualitative behavior of complex materials in finite-temperature conditions. 
Even though \emph{ad hoc} models trained for more restricted sets of compositions and thermodynamic conditions, \rev{and the incorporation of magnetic effects at least at the level of spin-polarized DFT,}  are still needed to achieve errors of a few meV/atom, the semi-quantitative accuracy we achieve is sufficient to gather insights on subtle phenomena such as surface segregation, to study in a detailed way the structure and energetics of specific alloy compositions, and to propose new alloy compositions for applications to heterogeneous catalysis.

\section*{Data availability}

All data and code used to train the HEA25S-4-NN model, as well as the fitted parameters and code to run the simulations discussed in this work is available in the \SM{} or from publicly-accessible repositories.

\section*{Acknowedgements}

MC, AM and NL acknowledge support from the NCCR MARVEL, funded by the Swiss National Science Foundation (grant number 182892) and from an Industrial Grant from BASF. GF acknowledges support by the Swiss Platform for Advanced Scientific Computing (PASC). Electronic-structure calculations were performed within the scope of a CSCS project (ID: s1092).


\begin{thebibliography}{82}%
\makeatletter
\providecommand \@ifxundefined [1]{%
 \@ifx{#1\undefined}
}%
\providecommand \@ifnum [1]{%
 \ifnum #1\expandafter \@firstoftwo
 \else \expandafter \@secondoftwo
 \fi
}%
\providecommand \@ifx [1]{%
 \ifx #1\expandafter \@firstoftwo
 \else \expandafter \@secondoftwo
 \fi
}%
\providecommand \natexlab [1]{#1}%
\providecommand \enquote  [1]{``#1''}%
\providecommand \bibnamefont  [1]{#1}%
\providecommand \bibfnamefont [1]{#1}%
\providecommand \citenamefont [1]{#1}%
\providecommand \href@noop [0]{\@secondoftwo}%
\providecommand \href [0]{\begingroup \@sanitize@url \@href}%
\providecommand \@href[1]{\@@startlink{#1}\@@href}%
\providecommand \@@href[1]{\endgroup#1\@@endlink}%
\providecommand \@sanitize@url [0]{\catcode `\\12\catcode `\$12\catcode
  `\&12\catcode `\#12\catcode `\^12\catcode `\_12\catcode `\%12\relax}%
\providecommand \@@startlink[1]{}%
\providecommand \@@endlink[0]{}%
\providecommand \url  [0]{\begingroup\@sanitize@url \@url }%
\providecommand \@url [1]{\endgroup\@href {#1}{\urlprefix }}%
\providecommand \urlprefix  [0]{URL }%
\providecommand \Eprint [0]{\href }%
\providecommand \doibase [0]{http://dx.doi.org/}%
\providecommand \selectlanguage [0]{\@gobble}%
\providecommand \bibinfo  [0]{\@secondoftwo}%
\providecommand \bibfield  [0]{\@secondoftwo}%
\providecommand \translation [1]{[#1]}%
\providecommand \BibitemOpen [0]{}%
\providecommand \bibitemStop [0]{}%
\providecommand \bibitemNoStop [0]{.\EOS\space}%
\providecommand \EOS [0]{\spacefactor3000\relax}%
\providecommand \BibitemShut  [1]{\csname bibitem#1\endcsname}%
\let\auto@bib@innerbib\@empty
\bibitem [{\citenamefont {Yeh}\ \emph {et~al.}(2004)\citenamefont {Yeh},
  \citenamefont {Chen}, \citenamefont {Lin}, \citenamefont {Gan}, \citenamefont
  {Chin}, \citenamefont {Shun}, \citenamefont {Tsau},\ and\ \citenamefont
  {Chang}}]{yeh+04aem}%
  \BibitemOpen
  \bibfield  {author} {\bibinfo {author} {\bibfnamefont {J.-W.}\ \bibnamefont
  {Yeh}}, \bibinfo {author} {\bibfnamefont {S.-K.}\ \bibnamefont {Chen}},
  \bibinfo {author} {\bibfnamefont {S.-J.}\ \bibnamefont {Lin}}, \bibinfo
  {author} {\bibfnamefont {J.-Y.}\ \bibnamefont {Gan}}, \bibinfo {author}
  {\bibfnamefont {T.-S.}\ \bibnamefont {Chin}}, \bibinfo {author}
  {\bibfnamefont {T.-T.}\ \bibnamefont {Shun}}, \bibinfo {author}
  {\bibfnamefont {C.-H.}\ \bibnamefont {Tsau}}, \ and\ \bibinfo {author}
  {\bibfnamefont {S.-Y.}\ \bibnamefont {Chang}},\ }\href {\doibase
  10.1002/adem.200300567} {\bibfield  {journal} {\bibinfo  {journal} {Adv. Eng.
  Mater.}\ }\textbf {\bibinfo {volume} {6}},\ \bibinfo {pages} {299} (\bibinfo
  {year} {2004})}\BibitemShut {NoStop}%
\bibitem [{\citenamefont {Cantor}\ \emph {et~al.}(2004)\citenamefont {Cantor},
  \citenamefont {Chang}, \citenamefont {Knight},\ and\ \citenamefont
  {Vincent}}]{cant+04msea}%
  \BibitemOpen
  \bibfield  {author} {\bibinfo {author} {\bibfnamefont {B.}~\bibnamefont
  {Cantor}}, \bibinfo {author} {\bibfnamefont {I.}~\bibnamefont {Chang}},
  \bibinfo {author} {\bibfnamefont {P.}~\bibnamefont {Knight}}, \ and\ \bibinfo
  {author} {\bibfnamefont {A.}~\bibnamefont {Vincent}},\ }\href {\doibase
  10.1016/j.msea.2003.10.257} {\bibfield  {journal} {\bibinfo  {journal}
  {Materials Science and Engineering: A}\ }\textbf {\bibinfo {volume}
  {375--377}},\ \bibinfo {pages} {213} (\bibinfo {year} {2004})}\BibitemShut
  {NoStop}%
\bibitem [{\citenamefont {Sun}\ and\ \citenamefont {Dai}(2021)}]{Sun2021}%
  \BibitemOpen
  \bibfield  {author} {\bibinfo {author} {\bibfnamefont {Y.}~\bibnamefont
  {Sun}}\ and\ \bibinfo {author} {\bibfnamefont {S.}~\bibnamefont {Dai}},\
  }\href {\doibase 10.1126/sciadv.abg1600} {\bibfield  {journal} {\bibinfo
  {journal} {Science Advances}\ }\textbf {\bibinfo {volume} {7}} (\bibinfo
  {year} {2021}),\ 10.1126/sciadv.abg1600}\BibitemShut {NoStop}%
\bibitem [{\citenamefont {Wang}\ \emph
  {et~al.}(2021{\natexlab{a}})\citenamefont {Wang}, \citenamefont {Yao},
  \citenamefont {Yu}, \citenamefont {Wang}, \citenamefont {Wu},\ and\
  \citenamefont {Zou}}]{Wang2021}%
  \BibitemOpen
  \bibfield  {author} {\bibinfo {author} {\bibfnamefont {B.}~\bibnamefont
  {Wang}}, \bibinfo {author} {\bibfnamefont {Y.}~\bibnamefont {Yao}}, \bibinfo
  {author} {\bibfnamefont {X.}~\bibnamefont {Yu}}, \bibinfo {author}
  {\bibfnamefont {C.}~\bibnamefont {Wang}}, \bibinfo {author} {\bibfnamefont
  {C.}~\bibnamefont {Wu}}, \ and\ \bibinfo {author} {\bibfnamefont
  {Z.}~\bibnamefont {Zou}},\ }\href {\doibase 10.1039/d1ta02718b} {\bibfield
  {journal} {\bibinfo  {journal} {Journal of Materials Chemistry A}\ }\textbf
  {\bibinfo {volume} {9}},\ \bibinfo {pages} {19410} (\bibinfo {year}
  {2021}{\natexlab{a}})}\BibitemShut {NoStop}%
\bibitem [{\citenamefont {{Kumar Katiyar}}\ \emph {et~al.}(2021)\citenamefont
  {{Kumar Katiyar}}, \citenamefont {Biswas}, \citenamefont {Yeh}, \citenamefont
  {Sharma},\ and\ \citenamefont {{Sekhar Tiwary}}}]{Katiyar2021}%
  \BibitemOpen
  \bibfield  {author} {\bibinfo {author} {\bibfnamefont {N.}~\bibnamefont
  {{Kumar Katiyar}}}, \bibinfo {author} {\bibfnamefont {K.}~\bibnamefont
  {Biswas}}, \bibinfo {author} {\bibfnamefont {J.-W.}\ \bibnamefont {Yeh}},
  \bibinfo {author} {\bibfnamefont {S.}~\bibnamefont {Sharma}}, \ and\ \bibinfo
  {author} {\bibfnamefont {C.}~\bibnamefont {{Sekhar Tiwary}}},\ }\href
  {\doibase https://doi.org/10.1016/j.nanoen.2021.106261} {\bibfield  {journal}
  {\bibinfo  {journal} {Nano Energy}\ }\textbf {\bibinfo {volume} {88}},\
  \bibinfo {pages} {106261} (\bibinfo {year} {2021})}\BibitemShut {NoStop}%
\bibitem [{\citenamefont {Xin}\ \emph {et~al.}(2020)\citenamefont {Xin},
  \citenamefont {Li}, \citenamefont {Qian}, \citenamefont {Zhu}, \citenamefont
  {Yuan}, \citenamefont {Jiang}, \citenamefont {Guo},\ and\ \citenamefont
  {Wang}}]{Xin2020}%
  \BibitemOpen
  \bibfield  {author} {\bibinfo {author} {\bibfnamefont {Y.}~\bibnamefont
  {Xin}}, \bibinfo {author} {\bibfnamefont {S.}~\bibnamefont {Li}}, \bibinfo
  {author} {\bibfnamefont {Y.}~\bibnamefont {Qian}}, \bibinfo {author}
  {\bibfnamefont {W.}~\bibnamefont {Zhu}}, \bibinfo {author} {\bibfnamefont
  {H.}~\bibnamefont {Yuan}}, \bibinfo {author} {\bibfnamefont {P.}~\bibnamefont
  {Jiang}}, \bibinfo {author} {\bibfnamefont {R.}~\bibnamefont {Guo}}, \ and\
  \bibinfo {author} {\bibfnamefont {L.}~\bibnamefont {Wang}},\ }\href {\doibase
  10.1021/acscatal.0c03617} {\bibfield  {journal} {\bibinfo  {journal} {ACS
  Catalysis}\ }\textbf {\bibinfo {volume} {10}},\ \bibinfo {pages} {11280}
  (\bibinfo {year} {2020})},\ \Eprint
  {http://arxiv.org/abs/https://doi.org/10.1021/acscatal.0c03617}
  {https://doi.org/10.1021/acscatal.0c03617} \BibitemShut {NoStop}%
\bibitem [{\citenamefont {Yu}\ \emph {et~al.}(2022)\citenamefont {Yu},
  \citenamefont {Zeng}, \citenamefont {Li}, \citenamefont {Lin}, \citenamefont
  {Liu}, \citenamefont {Shi}, \citenamefont {Qiu}, \citenamefont {Yuan},\ and\
  \citenamefont {Yao}}]{Yu2022}%
  \BibitemOpen
  \bibfield  {author} {\bibinfo {author} {\bibfnamefont {L.}~\bibnamefont
  {Yu}}, \bibinfo {author} {\bibfnamefont {K.}~\bibnamefont {Zeng}}, \bibinfo
  {author} {\bibfnamefont {C.}~\bibnamefont {Li}}, \bibinfo {author}
  {\bibfnamefont {X.}~\bibnamefont {Lin}}, \bibinfo {author} {\bibfnamefont
  {H.}~\bibnamefont {Liu}}, \bibinfo {author} {\bibfnamefont {W.}~\bibnamefont
  {Shi}}, \bibinfo {author} {\bibfnamefont {H.-J.}\ \bibnamefont {Qiu}},
  \bibinfo {author} {\bibfnamefont {Y.}~\bibnamefont {Yuan}}, \ and\ \bibinfo
  {author} {\bibfnamefont {Y.}~\bibnamefont {Yao}},\ }\href {\doibase
  https://doi.org/10.1002/cey2.228} {\bibfield  {journal} {\bibinfo  {journal}
  {Carbon Energy}\ }\textbf {\bibinfo {volume} {4}},\ \bibinfo {pages} {731}
  (\bibinfo {year} {2022})},\ \Eprint
  {http://arxiv.org/abs/https://onlinelibrary.wiley.com/doi/pdf/10.1002/cey2.228}
  {https://onlinelibrary.wiley.com/doi/pdf/10.1002/cey2.228} \BibitemShut
  {NoStop}%
\bibitem [{\citenamefont {Huo}\ \emph {et~al.}(2021)\citenamefont {Huo},
  \citenamefont {Wang}, \citenamefont {Zhu}, \citenamefont {Zhang},
  \citenamefont {Fang}, \citenamefont {Xie},\ and\ \citenamefont
  {Jiang}}]{Huo2021}%
  \BibitemOpen
  \bibfield  {author} {\bibinfo {author} {\bibfnamefont {W.-Y.}\ \bibnamefont
  {Huo}}, \bibinfo {author} {\bibfnamefont {S.-Q.}\ \bibnamefont {Wang}},
  \bibinfo {author} {\bibfnamefont {W.-H.}\ \bibnamefont {Zhu}}, \bibinfo
  {author} {\bibfnamefont {Z.-L.}\ \bibnamefont {Zhang}}, \bibinfo {author}
  {\bibfnamefont {F.}~\bibnamefont {Fang}}, \bibinfo {author} {\bibfnamefont
  {Z.-H.}\ \bibnamefont {Xie}}, \ and\ \bibinfo {author} {\bibfnamefont
  {J.-Q.}\ \bibnamefont {Jiang}},\ }\href {\doibase 10.1007/s42864-021-00084-8}
  {\bibfield  {journal} {\bibinfo  {journal} {Tungsten}\ }\textbf {\bibinfo
  {volume} {3}},\ \bibinfo {pages} {161} (\bibinfo {year} {2021})}\BibitemShut
  {NoStop}%
\bibitem [{\citenamefont {Zhang}\ \emph {et~al.}(2021)\citenamefont {Zhang},
  \citenamefont {Wang},\ and\ \citenamefont {Wang}}]{Zhang2021}%
  \BibitemOpen
  \bibfield  {author} {\bibinfo {author} {\bibfnamefont {Y.}~\bibnamefont
  {Zhang}}, \bibinfo {author} {\bibfnamefont {D.}~\bibnamefont {Wang}}, \ and\
  \bibinfo {author} {\bibfnamefont {S.}~\bibnamefont {Wang}},\ }\href {\doibase
  10.1002/smll.202104339} {\bibfield  {journal} {\bibinfo  {journal} {Small}\
  }\textbf {\bibinfo {volume} {18}},\ \bibinfo {pages} {2104339} (\bibinfo
  {year} {2021})}\BibitemShut {NoStop}%
\bibitem [{\citenamefont {Huo}\ \emph {et~al.}(2022)\citenamefont {Huo},
  \citenamefont {Yu}, \citenamefont {Xing}, \citenamefont {Zuo},\ and\
  \citenamefont {Zhang}}]{Huo2022}%
  \BibitemOpen
  \bibfield  {author} {\bibinfo {author} {\bibfnamefont {X.}~\bibnamefont
  {Huo}}, \bibinfo {author} {\bibfnamefont {H.}~\bibnamefont {Yu}}, \bibinfo
  {author} {\bibfnamefont {B.}~\bibnamefont {Xing}}, \bibinfo {author}
  {\bibfnamefont {X.}~\bibnamefont {Zuo}}, \ and\ \bibinfo {author}
  {\bibfnamefont {N.}~\bibnamefont {Zhang}},\ }\href {\doibase
  10.1002/tcr.202200175} {\bibfield  {journal} {\bibinfo  {journal} {The
  Chemical Record}\ } (\bibinfo {year} {2022}),\
  10.1002/tcr.202200175}\BibitemShut {NoStop}%
\bibitem [{\citenamefont {Zhang}\ \emph {et~al.}(2018)\citenamefont {Zhang},
  \citenamefont {Ming}, \citenamefont {Kang}, \citenamefont {Huang},
  \citenamefont {Zhang}, \citenamefont {Zheng},\ and\ \citenamefont
  {Bi}}]{Zhang2018}%
  \BibitemOpen
  \bibfield  {author} {\bibinfo {author} {\bibfnamefont {G.}~\bibnamefont
  {Zhang}}, \bibinfo {author} {\bibfnamefont {K.}~\bibnamefont {Ming}},
  \bibinfo {author} {\bibfnamefont {J.}~\bibnamefont {Kang}}, \bibinfo {author}
  {\bibfnamefont {Q.}~\bibnamefont {Huang}}, \bibinfo {author} {\bibfnamefont
  {Z.}~\bibnamefont {Zhang}}, \bibinfo {author} {\bibfnamefont
  {X.}~\bibnamefont {Zheng}}, \ and\ \bibinfo {author} {\bibfnamefont
  {X.}~\bibnamefont {Bi}},\ }\href {\doibase 10.1016/j.electacta.2018.05.035}
  {\bibfield  {journal} {\bibinfo  {journal} {Electrochimica Acta}\ }\textbf
  {\bibinfo {volume} {279}},\ \bibinfo {pages} {19} (\bibinfo {year}
  {2018})}\BibitemShut {NoStop}%
\bibitem [{\citenamefont {Bondesgaard}\ \emph {et~al.}(2019)\citenamefont
  {Bondesgaard}, \citenamefont {Broge}, \citenamefont {Mamakhel}, \citenamefont
  {Bremholm},\ and\ \citenamefont {Iversen}}]{Bondesgaard2019}%
  \BibitemOpen
  \bibfield  {author} {\bibinfo {author} {\bibfnamefont {M.}~\bibnamefont
  {Bondesgaard}}, \bibinfo {author} {\bibfnamefont {N.~L.~N.}\ \bibnamefont
  {Broge}}, \bibinfo {author} {\bibfnamefont {A.}~\bibnamefont {Mamakhel}},
  \bibinfo {author} {\bibfnamefont {M.}~\bibnamefont {Bremholm}}, \ and\
  \bibinfo {author} {\bibfnamefont {B.~B.}\ \bibnamefont {Iversen}},\ }\href
  {\doibase 10.1002/adfm.201905933} {\bibfield  {journal} {\bibinfo  {journal}
  {Advanced Functional Materials}\ }\textbf {\bibinfo {volume} {29}},\ \bibinfo
  {pages} {1905933} (\bibinfo {year} {2019})}\BibitemShut {NoStop}%
\bibitem [{\citenamefont {Glasscott}\ \emph {et~al.}(2019)\citenamefont
  {Glasscott}, \citenamefont {Pendergast}, \citenamefont {Goines},
  \citenamefont {Bishop}, \citenamefont {Hoang}, \citenamefont {Renault},\ and\
  \citenamefont {Dick}}]{Glasscott2019}%
  \BibitemOpen
  \bibfield  {author} {\bibinfo {author} {\bibfnamefont {M.~W.}\ \bibnamefont
  {Glasscott}}, \bibinfo {author} {\bibfnamefont {A.~D.}\ \bibnamefont
  {Pendergast}}, \bibinfo {author} {\bibfnamefont {S.}~\bibnamefont {Goines}},
  \bibinfo {author} {\bibfnamefont {A.~R.}\ \bibnamefont {Bishop}}, \bibinfo
  {author} {\bibfnamefont {A.~T.}\ \bibnamefont {Hoang}}, \bibinfo {author}
  {\bibfnamefont {C.}~\bibnamefont {Renault}}, \ and\ \bibinfo {author}
  {\bibfnamefont {J.~E.}\ \bibnamefont {Dick}},\ }\href {\doibase
  10.1038/s41467-019-10303-z} {\bibfield  {journal} {\bibinfo  {journal}
  {Nature Communications}\ }\textbf {\bibinfo {volume} {10}} (\bibinfo {year}
  {2019}),\ 10.1038/s41467-019-10303-z}\BibitemShut {NoStop}%
\bibitem [{\citenamefont {Jin}\ \emph {et~al.}(2019)\citenamefont {Jin},
  \citenamefont {Lv}, \citenamefont {Jia}, \citenamefont {Liu}, \citenamefont
  {Li}, \citenamefont {Chen}, \citenamefont {Lin}, \citenamefont {Xie},
  \citenamefont {Liu}, \citenamefont {Sun},\ and\ \citenamefont
  {Qiu}}]{Jin2019}%
  \BibitemOpen
  \bibfield  {author} {\bibinfo {author} {\bibfnamefont {Z.}~\bibnamefont
  {Jin}}, \bibinfo {author} {\bibfnamefont {J.}~\bibnamefont {Lv}}, \bibinfo
  {author} {\bibfnamefont {H.}~\bibnamefont {Jia}}, \bibinfo {author}
  {\bibfnamefont {W.}~\bibnamefont {Liu}}, \bibinfo {author} {\bibfnamefont
  {H.}~\bibnamefont {Li}}, \bibinfo {author} {\bibfnamefont {Z.}~\bibnamefont
  {Chen}}, \bibinfo {author} {\bibfnamefont {X.}~\bibnamefont {Lin}}, \bibinfo
  {author} {\bibfnamefont {G.}~\bibnamefont {Xie}}, \bibinfo {author}
  {\bibfnamefont {X.}~\bibnamefont {Liu}}, \bibinfo {author} {\bibfnamefont
  {S.}~\bibnamefont {Sun}}, \ and\ \bibinfo {author} {\bibfnamefont {H.-J.}\
  \bibnamefont {Qiu}},\ }\href {\doibase 10.1002/smll.201904180} {\bibfield
  {journal} {\bibinfo  {journal} {Small}\ }\textbf {\bibinfo {volume} {15}},\
  \bibinfo {pages} {1904180} (\bibinfo {year} {2019})}\BibitemShut {NoStop}%
\bibitem [{\citenamefont {Lacey}\ \emph {et~al.}(2019)\citenamefont {Lacey},
  \citenamefont {Dong}, \citenamefont {Huang}, \citenamefont {Luo},
  \citenamefont {Xie}, \citenamefont {Lin}, \citenamefont {Kirsch},
  \citenamefont {Vattipalli}, \citenamefont {Povinelli}, \citenamefont {Fan},
  \citenamefont {Shahbazian-Yassar}, \citenamefont {Wang},\ and\ \citenamefont
  {Hu}}]{Lacey2019}%
  \BibitemOpen
  \bibfield  {author} {\bibinfo {author} {\bibfnamefont {S.~D.}\ \bibnamefont
  {Lacey}}, \bibinfo {author} {\bibfnamefont {Q.}~\bibnamefont {Dong}},
  \bibinfo {author} {\bibfnamefont {Z.}~\bibnamefont {Huang}}, \bibinfo
  {author} {\bibfnamefont {J.}~\bibnamefont {Luo}}, \bibinfo {author}
  {\bibfnamefont {H.}~\bibnamefont {Xie}}, \bibinfo {author} {\bibfnamefont
  {Z.}~\bibnamefont {Lin}}, \bibinfo {author} {\bibfnamefont {D.~J.}\
  \bibnamefont {Kirsch}}, \bibinfo {author} {\bibfnamefont {V.}~\bibnamefont
  {Vattipalli}}, \bibinfo {author} {\bibfnamefont {C.}~\bibnamefont
  {Povinelli}}, \bibinfo {author} {\bibfnamefont {W.}~\bibnamefont {Fan}},
  \bibinfo {author} {\bibfnamefont {R.}~\bibnamefont {Shahbazian-Yassar}},
  \bibinfo {author} {\bibfnamefont {D.}~\bibnamefont {Wang}}, \ and\ \bibinfo
  {author} {\bibfnamefont {L.}~\bibnamefont {Hu}},\ }\href {\doibase
  10.1021/acs.nanolett.9b01523} {\bibfield  {journal} {\bibinfo  {journal}
  {Nano Letters}\ }\textbf {\bibinfo {volume} {19}},\ \bibinfo {pages} {5149}
  (\bibinfo {year} {2019})}\BibitemShut {NoStop}%
\bibitem [{\citenamefont {Liu}\ \emph {et~al.}(2019)\citenamefont {Liu},
  \citenamefont {Zhang}, \citenamefont {Okejiri}, \citenamefont {Yang},
  \citenamefont {Zhou},\ and\ \citenamefont {Dai}}]{Liu2019}%
  \BibitemOpen
  \bibfield  {author} {\bibinfo {author} {\bibfnamefont {M.}~\bibnamefont
  {Liu}}, \bibinfo {author} {\bibfnamefont {Z.}~\bibnamefont {Zhang}}, \bibinfo
  {author} {\bibfnamefont {F.}~\bibnamefont {Okejiri}}, \bibinfo {author}
  {\bibfnamefont {S.}~\bibnamefont {Yang}}, \bibinfo {author} {\bibfnamefont
  {S.}~\bibnamefont {Zhou}}, \ and\ \bibinfo {author} {\bibfnamefont
  {S.}~\bibnamefont {Dai}},\ }\href {\doibase 10.1002/admi.201900015}
  {\bibfield  {journal} {\bibinfo  {journal} {Advanced Materials Interfaces}\
  }\textbf {\bibinfo {volume} {6}},\ \bibinfo {pages} {1900015} (\bibinfo
  {year} {2019})}\BibitemShut {NoStop}%
\bibitem [{\citenamefont {Qiu}\ \emph {et~al.}(2019{\natexlab{a}})\citenamefont
  {Qiu}, \citenamefont {Fang}, \citenamefont {Gao}, \citenamefont {Wen},
  \citenamefont {Lv}, \citenamefont {Li}, \citenamefont {Xie}, \citenamefont
  {Liu},\ and\ \citenamefont {Sun}}]{Qiu2019}%
  \BibitemOpen
  \bibfield  {author} {\bibinfo {author} {\bibfnamefont {H.-J.}\ \bibnamefont
  {Qiu}}, \bibinfo {author} {\bibfnamefont {G.}~\bibnamefont {Fang}}, \bibinfo
  {author} {\bibfnamefont {J.}~\bibnamefont {Gao}}, \bibinfo {author}
  {\bibfnamefont {Y.}~\bibnamefont {Wen}}, \bibinfo {author} {\bibfnamefont
  {J.}~\bibnamefont {Lv}}, \bibinfo {author} {\bibfnamefont {H.}~\bibnamefont
  {Li}}, \bibinfo {author} {\bibfnamefont {G.}~\bibnamefont {Xie}}, \bibinfo
  {author} {\bibfnamefont {X.}~\bibnamefont {Liu}}, \ and\ \bibinfo {author}
  {\bibfnamefont {S.}~\bibnamefont {Sun}},\ }\href {\doibase
  10.1021/acsmaterialslett.9b00414} {\bibfield  {journal} {\bibinfo  {journal}
  {{ACS} Materials Letters}\ }\textbf {\bibinfo {volume} {1}},\ \bibinfo
  {pages} {526} (\bibinfo {year} {2019}{\natexlab{a}})}\BibitemShut {NoStop}%
\bibitem [{\citenamefont {Qiu}\ \emph {et~al.}(2019{\natexlab{b}})\citenamefont
  {Qiu}, \citenamefont {Fang}, \citenamefont {Wen}, \citenamefont {Liu},
  \citenamefont {Xie}, \citenamefont {Liu},\ and\ \citenamefont
  {Sun}}]{Qiu2019a}%
  \BibitemOpen
  \bibfield  {author} {\bibinfo {author} {\bibfnamefont {H.-J.}\ \bibnamefont
  {Qiu}}, \bibinfo {author} {\bibfnamefont {G.}~\bibnamefont {Fang}}, \bibinfo
  {author} {\bibfnamefont {Y.}~\bibnamefont {Wen}}, \bibinfo {author}
  {\bibfnamefont {P.}~\bibnamefont {Liu}}, \bibinfo {author} {\bibfnamefont
  {G.}~\bibnamefont {Xie}}, \bibinfo {author} {\bibfnamefont {X.}~\bibnamefont
  {Liu}}, \ and\ \bibinfo {author} {\bibfnamefont {S.}~\bibnamefont {Sun}},\
  }\href {\doibase 10.1039/c9ta00505f} {\bibfield  {journal} {\bibinfo
  {journal} {Journal of Materials Chemistry A}\ }\textbf {\bibinfo {volume}
  {7}},\ \bibinfo {pages} {6499} (\bibinfo {year}
  {2019}{\natexlab{b}})}\BibitemShut {NoStop}%
\bibitem [{\citenamefont {Gao}\ \emph {et~al.}(2020)\citenamefont {Gao},
  \citenamefont {Hao}, \citenamefont {Huang}, \citenamefont {Yuan},
  \citenamefont {Han}, \citenamefont {Lei}, \citenamefont {Zhang},
  \citenamefont {Shahbazian-Yassar},\ and\ \citenamefont {Lu}}]{Gao2020}%
  \BibitemOpen
  \bibfield  {author} {\bibinfo {author} {\bibfnamefont {S.}~\bibnamefont
  {Gao}}, \bibinfo {author} {\bibfnamefont {S.}~\bibnamefont {Hao}}, \bibinfo
  {author} {\bibfnamefont {Z.}~\bibnamefont {Huang}}, \bibinfo {author}
  {\bibfnamefont {Y.}~\bibnamefont {Yuan}}, \bibinfo {author} {\bibfnamefont
  {S.}~\bibnamefont {Han}}, \bibinfo {author} {\bibfnamefont {L.}~\bibnamefont
  {Lei}}, \bibinfo {author} {\bibfnamefont {X.}~\bibnamefont {Zhang}}, \bibinfo
  {author} {\bibfnamefont {R.}~\bibnamefont {Shahbazian-Yassar}}, \ and\
  \bibinfo {author} {\bibfnamefont {J.}~\bibnamefont {Lu}},\ }\href {\doibase
  10.1038/s41467-020-15934-1} {\bibfield  {journal} {\bibinfo  {journal}
  {Nature Communications}\ }\textbf {\bibinfo {volume} {11}} (\bibinfo {year}
  {2020}),\ 10.1038/s41467-020-15934-1}\BibitemShut {NoStop}%
\bibitem [{\citenamefont {Huang}\ \emph {et~al.}(2020)\citenamefont {Huang},
  \citenamefont {Zhang}, \citenamefont {Wu}, \citenamefont {Zhang},
  \citenamefont {Peng}, \citenamefont {Cao}, \citenamefont {Zhang},
  \citenamefont {Li},\ and\ \citenamefont {Huang}}]{Huang2020}%
  \BibitemOpen
  \bibfield  {author} {\bibinfo {author} {\bibfnamefont {K.}~\bibnamefont
  {Huang}}, \bibinfo {author} {\bibfnamefont {B.}~\bibnamefont {Zhang}},
  \bibinfo {author} {\bibfnamefont {J.}~\bibnamefont {Wu}}, \bibinfo {author}
  {\bibfnamefont {T.}~\bibnamefont {Zhang}}, \bibinfo {author} {\bibfnamefont
  {D.}~\bibnamefont {Peng}}, \bibinfo {author} {\bibfnamefont {X.}~\bibnamefont
  {Cao}}, \bibinfo {author} {\bibfnamefont {Z.}~\bibnamefont {Zhang}}, \bibinfo
  {author} {\bibfnamefont {Z.}~\bibnamefont {Li}}, \ and\ \bibinfo {author}
  {\bibfnamefont {Y.}~\bibnamefont {Huang}},\ }\href {\doibase
  10.1039/d0ta02125c} {\bibfield  {journal} {\bibinfo  {journal} {Journal of
  Materials Chemistry A}\ }\textbf {\bibinfo {volume} {8}},\ \bibinfo {pages}
  {11938} (\bibinfo {year} {2020})}\BibitemShut {NoStop}%
\bibitem [{\citenamefont {Wu}\ \emph {et~al.}(2020)\citenamefont {Wu},
  \citenamefont {Kusada}, \citenamefont {Yamamoto}, \citenamefont {Toriyama},
  \citenamefont {Matsumura}, \citenamefont {Gueye}, \citenamefont {Seo},
  \citenamefont {Kim}, \citenamefont {Hiroi}, \citenamefont {Sakata},
  \citenamefont {Kawaguchi}, \citenamefont {Kubota},\ and\ \citenamefont
  {Kitagawa}}]{Wu2020}%
  \BibitemOpen
  \bibfield  {author} {\bibinfo {author} {\bibfnamefont {D.}~\bibnamefont
  {Wu}}, \bibinfo {author} {\bibfnamefont {K.}~\bibnamefont {Kusada}}, \bibinfo
  {author} {\bibfnamefont {T.}~\bibnamefont {Yamamoto}}, \bibinfo {author}
  {\bibfnamefont {T.}~\bibnamefont {Toriyama}}, \bibinfo {author}
  {\bibfnamefont {S.}~\bibnamefont {Matsumura}}, \bibinfo {author}
  {\bibfnamefont {I.}~\bibnamefont {Gueye}}, \bibinfo {author} {\bibfnamefont
  {O.}~\bibnamefont {Seo}}, \bibinfo {author} {\bibfnamefont {J.}~\bibnamefont
  {Kim}}, \bibinfo {author} {\bibfnamefont {S.}~\bibnamefont {Hiroi}}, \bibinfo
  {author} {\bibfnamefont {O.}~\bibnamefont {Sakata}}, \bibinfo {author}
  {\bibfnamefont {S.}~\bibnamefont {Kawaguchi}}, \bibinfo {author}
  {\bibfnamefont {Y.}~\bibnamefont {Kubota}}, \ and\ \bibinfo {author}
  {\bibfnamefont {H.}~\bibnamefont {Kitagawa}},\ }\href {\doibase
  10.1039/d0sc02351e} {\bibfield  {journal} {\bibinfo  {journal} {Chemical
  Science}\ }\textbf {\bibinfo {volume} {11}},\ \bibinfo {pages} {12731}
  (\bibinfo {year} {2020})}\BibitemShut {NoStop}%
\bibitem [{\citenamefont {Katiyar}\ \emph {et~al.}(2020)\citenamefont
  {Katiyar}, \citenamefont {Nellaiappan}, \citenamefont {Kumar}, \citenamefont
  {Malviya}, \citenamefont {Pradeep}, \citenamefont {Singh}, \citenamefont
  {Sharma}, \citenamefont {Tiwary},\ and\ \citenamefont
  {Biswas}}]{Katiyar2020}%
  \BibitemOpen
  \bibfield  {author} {\bibinfo {author} {\bibfnamefont {N.~K.}\ \bibnamefont
  {Katiyar}}, \bibinfo {author} {\bibfnamefont {S.}~\bibnamefont
  {Nellaiappan}}, \bibinfo {author} {\bibfnamefont {R.}~\bibnamefont {Kumar}},
  \bibinfo {author} {\bibfnamefont {K.~D.}\ \bibnamefont {Malviya}}, \bibinfo
  {author} {\bibfnamefont {K.}~\bibnamefont {Pradeep}}, \bibinfo {author}
  {\bibfnamefont {A.~K.}\ \bibnamefont {Singh}}, \bibinfo {author}
  {\bibfnamefont {S.}~\bibnamefont {Sharma}}, \bibinfo {author} {\bibfnamefont
  {C.~S.}\ \bibnamefont {Tiwary}}, \ and\ \bibinfo {author} {\bibfnamefont
  {K.}~\bibnamefont {Biswas}},\ }\href {\doibase
  https://doi.org/10.1016/j.mtener.2020.100393} {\bibfield  {journal} {\bibinfo
   {journal} {Materials Today Energy}\ }\textbf {\bibinfo {volume} {16}},\
  \bibinfo {pages} {100393} (\bibinfo {year} {2020})}\BibitemShut {NoStop}%
\bibitem [{\citenamefont {Chen}\ \emph {et~al.}(2015)\citenamefont {Chen},
  \citenamefont {Si}, \citenamefont {Gao}, \citenamefont {Frenzel},
  \citenamefont {Sun}, \citenamefont {Eggeler},\ and\ \citenamefont
  {Zhang}}]{Chen2015}%
  \BibitemOpen
  \bibfield  {author} {\bibinfo {author} {\bibfnamefont {X.}~\bibnamefont
  {Chen}}, \bibinfo {author} {\bibfnamefont {C.}~\bibnamefont {Si}}, \bibinfo
  {author} {\bibfnamefont {Y.}~\bibnamefont {Gao}}, \bibinfo {author}
  {\bibfnamefont {J.}~\bibnamefont {Frenzel}}, \bibinfo {author} {\bibfnamefont
  {J.}~\bibnamefont {Sun}}, \bibinfo {author} {\bibfnamefont {G.}~\bibnamefont
  {Eggeler}}, \ and\ \bibinfo {author} {\bibfnamefont {Z.}~\bibnamefont
  {Zhang}},\ }\href {\doibase 10.1016/j.jpowsour.2014.09.076} {\bibfield
  {journal} {\bibinfo  {journal} {Journal of Power Sources}\ }\textbf {\bibinfo
  {volume} {273}},\ \bibinfo {pages} {324} (\bibinfo {year}
  {2015})}\BibitemShut {NoStop}%
\bibitem [{\citenamefont {L\"{o}ffler}\ \emph {et~al.}(2018)\citenamefont
  {L\"{o}ffler}, \citenamefont {Meyer}, \citenamefont {Savan}, \citenamefont
  {Wilde}, \citenamefont {Manj{\'{o}}n}, \citenamefont {Chen}, \citenamefont
  {Ventosa}, \citenamefont {Scheu}, \citenamefont {Ludwig},\ and\ \citenamefont
  {Schuhmann}}]{Lffler2018}%
  \BibitemOpen
  \bibfield  {author} {\bibinfo {author} {\bibfnamefont {T.}~\bibnamefont
  {L\"{o}ffler}}, \bibinfo {author} {\bibfnamefont {H.}~\bibnamefont {Meyer}},
  \bibinfo {author} {\bibfnamefont {A.}~\bibnamefont {Savan}}, \bibinfo
  {author} {\bibfnamefont {P.}~\bibnamefont {Wilde}}, \bibinfo {author}
  {\bibfnamefont {A.~G.}\ \bibnamefont {Manj{\'{o}}n}}, \bibinfo {author}
  {\bibfnamefont {Y.-T.}\ \bibnamefont {Chen}}, \bibinfo {author}
  {\bibfnamefont {E.}~\bibnamefont {Ventosa}}, \bibinfo {author} {\bibfnamefont
  {C.}~\bibnamefont {Scheu}}, \bibinfo {author} {\bibfnamefont
  {A.}~\bibnamefont {Ludwig}}, \ and\ \bibinfo {author} {\bibfnamefont
  {W.}~\bibnamefont {Schuhmann}},\ }\href {\doibase 10.1002/aenm.201802269}
  {\bibfield  {journal} {\bibinfo  {journal} {Advanced Energy Materials}\
  }\textbf {\bibinfo {volume} {8}},\ \bibinfo {pages} {1802269} (\bibinfo
  {year} {2018})}\BibitemShut {NoStop}%
\bibitem [{\citenamefont {Li}\ \emph {et~al.}(2020)\citenamefont {Li},
  \citenamefont {Tang}, \citenamefont {Jia}, \citenamefont {Li}, \citenamefont
  {Xie}, \citenamefont {Liu}, \citenamefont {Lin},\ and\ \citenamefont
  {Qiu}}]{Li2020}%
  \BibitemOpen
  \bibfield  {author} {\bibinfo {author} {\bibfnamefont {S.}~\bibnamefont
  {Li}}, \bibinfo {author} {\bibfnamefont {X.}~\bibnamefont {Tang}}, \bibinfo
  {author} {\bibfnamefont {H.}~\bibnamefont {Jia}}, \bibinfo {author}
  {\bibfnamefont {H.}~\bibnamefont {Li}}, \bibinfo {author} {\bibfnamefont
  {G.}~\bibnamefont {Xie}}, \bibinfo {author} {\bibfnamefont {X.}~\bibnamefont
  {Liu}}, \bibinfo {author} {\bibfnamefont {X.}~\bibnamefont {Lin}}, \ and\
  \bibinfo {author} {\bibfnamefont {H.-J.}\ \bibnamefont {Qiu}},\ }\href
  {\doibase 10.1016/j.jcat.2020.01.024} {\bibfield  {journal} {\bibinfo
  {journal} {Journal of Catalysis}\ }\textbf {\bibinfo {volume} {383}},\
  \bibinfo {pages} {164} (\bibinfo {year} {2020})}\BibitemShut {NoStop}%
\bibitem [{\citenamefont {Pedersen}\ \emph
  {et~al.}(2021{\natexlab{a}})\citenamefont {Pedersen}, \citenamefont
  {Clausen}, \citenamefont {Krysiak}, \citenamefont {Xiao}, \citenamefont
  {Batchelor}, \citenamefont {L\"{o}ffler}, \citenamefont {Mints},
  \citenamefont {Banko}, \citenamefont {Arenz}, \citenamefont {Savan},
  \citenamefont {Schuhmann}, \citenamefont {Ludwig},\ and\ \citenamefont
  {Rossmeisl}}]{Pedersen2021}%
  \BibitemOpen
  \bibfield  {author} {\bibinfo {author} {\bibfnamefont {J.~K.}\ \bibnamefont
  {Pedersen}}, \bibinfo {author} {\bibfnamefont {C.~M.}\ \bibnamefont
  {Clausen}}, \bibinfo {author} {\bibfnamefont {O.~A.}\ \bibnamefont
  {Krysiak}}, \bibinfo {author} {\bibfnamefont {B.}~\bibnamefont {Xiao}},
  \bibinfo {author} {\bibfnamefont {T.~A.~A.}\ \bibnamefont {Batchelor}},
  \bibinfo {author} {\bibfnamefont {T.}~\bibnamefont {L\"{o}ffler}}, \bibinfo
  {author} {\bibfnamefont {V.~A.}\ \bibnamefont {Mints}}, \bibinfo {author}
  {\bibfnamefont {L.}~\bibnamefont {Banko}}, \bibinfo {author} {\bibfnamefont
  {M.}~\bibnamefont {Arenz}}, \bibinfo {author} {\bibfnamefont
  {A.}~\bibnamefont {Savan}}, \bibinfo {author} {\bibfnamefont
  {W.}~\bibnamefont {Schuhmann}}, \bibinfo {author} {\bibfnamefont
  {A.}~\bibnamefont {Ludwig}}, \ and\ \bibinfo {author} {\bibfnamefont
  {J.}~\bibnamefont {Rossmeisl}},\ }\href {\doibase 10.1002/ange.202108116}
  {\bibfield  {journal} {\bibinfo  {journal} {Angewandte Chemie}\ }\textbf
  {\bibinfo {volume} {133}},\ \bibinfo {pages} {24346} (\bibinfo {year}
  {2021}{\natexlab{a}})}\BibitemShut {NoStop}%
\bibitem [{\citenamefont {Barranco}\ and\ \citenamefont
  {Pierna}(2008)}]{Barranco2008}%
  \BibitemOpen
  \bibfield  {author} {\bibinfo {author} {\bibfnamefont {J.}~\bibnamefont
  {Barranco}}\ and\ \bibinfo {author} {\bibfnamefont {A.}~\bibnamefont
  {Pierna}},\ }\href {\doibase 10.1016/j.jnoncrysol.2008.04.053} {\bibfield
  {journal} {\bibinfo  {journal} {Journal of Non-Crystalline Solids}\ }\textbf
  {\bibinfo {volume} {354}},\ \bibinfo {pages} {5153} (\bibinfo {year}
  {2008})}\BibitemShut {NoStop}%
\bibitem [{\citenamefont {Tsai}\ \emph {et~al.}(2009)\citenamefont {Tsai},
  \citenamefont {Yeh}, \citenamefont {Wu}, \citenamefont {Hsieh},\ and\
  \citenamefont {Lin}}]{Tsai2009}%
  \BibitemOpen
  \bibfield  {author} {\bibinfo {author} {\bibfnamefont {C.-F.}\ \bibnamefont
  {Tsai}}, \bibinfo {author} {\bibfnamefont {K.-Y.}\ \bibnamefont {Yeh}},
  \bibinfo {author} {\bibfnamefont {P.-W.}\ \bibnamefont {Wu}}, \bibinfo
  {author} {\bibfnamefont {Y.-F.}\ \bibnamefont {Hsieh}}, \ and\ \bibinfo
  {author} {\bibfnamefont {P.}~\bibnamefont {Lin}},\ }\href {\doibase
  10.1016/j.jallcom.2008.12.055} {\bibfield  {journal} {\bibinfo  {journal}
  {Journal of Alloys and Compounds}\ }\textbf {\bibinfo {volume} {478}},\
  \bibinfo {pages} {868} (\bibinfo {year} {2009})}\BibitemShut {NoStop}%
\bibitem [{\citenamefont {Wang}\ \emph {et~al.}(2014)\citenamefont {Wang},
  \citenamefont {Wan}, \citenamefont {Xu}, \citenamefont {Tong},\ and\
  \citenamefont {Li}}]{Wang2014}%
  \BibitemOpen
  \bibfield  {author} {\bibinfo {author} {\bibfnamefont {A.-L.}\ \bibnamefont
  {Wang}}, \bibinfo {author} {\bibfnamefont {H.-C.}\ \bibnamefont {Wan}},
  \bibinfo {author} {\bibfnamefont {H.}~\bibnamefont {Xu}}, \bibinfo {author}
  {\bibfnamefont {Y.-X.}\ \bibnamefont {Tong}}, \ and\ \bibinfo {author}
  {\bibfnamefont {G.-R.}\ \bibnamefont {Li}},\ }\href {\doibase
  10.1016/j.electacta.2014.02.076} {\bibfield  {journal} {\bibinfo  {journal}
  {Electrochimica Acta}\ }\textbf {\bibinfo {volume} {127}},\ \bibinfo {pages}
  {448} (\bibinfo {year} {2014})}\BibitemShut {NoStop}%
\bibitem [{\citenamefont {Yusenko}\ \emph {et~al.}(2017)\citenamefont
  {Yusenko}, \citenamefont {Riva}, \citenamefont {Carvalho}, \citenamefont
  {Yusenko}, \citenamefont {Arnaboldi}, \citenamefont {Sukhikh}, \citenamefont
  {Hanfland},\ and\ \citenamefont {Gromilov}}]{Yusenko2017}%
  \BibitemOpen
  \bibfield  {author} {\bibinfo {author} {\bibfnamefont {K.~V.}\ \bibnamefont
  {Yusenko}}, \bibinfo {author} {\bibfnamefont {S.}~\bibnamefont {Riva}},
  \bibinfo {author} {\bibfnamefont {P.~A.}\ \bibnamefont {Carvalho}}, \bibinfo
  {author} {\bibfnamefont {M.~V.}\ \bibnamefont {Yusenko}}, \bibinfo {author}
  {\bibfnamefont {S.}~\bibnamefont {Arnaboldi}}, \bibinfo {author}
  {\bibfnamefont {A.~S.}\ \bibnamefont {Sukhikh}}, \bibinfo {author}
  {\bibfnamefont {M.}~\bibnamefont {Hanfland}}, \ and\ \bibinfo {author}
  {\bibfnamefont {S.~A.}\ \bibnamefont {Gromilov}},\ }\href {\doibase
  10.1016/j.scriptamat.2017.05.022} {\bibfield  {journal} {\bibinfo  {journal}
  {Scripta Materialia}\ }\textbf {\bibinfo {volume} {138}},\ \bibinfo {pages}
  {22} (\bibinfo {year} {2017})}\BibitemShut {NoStop}%
\bibitem [{\citenamefont {Pedersen}\ \emph {et~al.}(2020)\citenamefont
  {Pedersen}, \citenamefont {Batchelor}, \citenamefont {Bagger},\ and\
  \citenamefont {Rossmeisl}}]{Pedersen2020}%
  \BibitemOpen
  \bibfield  {author} {\bibinfo {author} {\bibfnamefont {J.~K.}\ \bibnamefont
  {Pedersen}}, \bibinfo {author} {\bibfnamefont {T.~A.~A.}\ \bibnamefont
  {Batchelor}}, \bibinfo {author} {\bibfnamefont {A.}~\bibnamefont {Bagger}}, \
  and\ \bibinfo {author} {\bibfnamefont {J.}~\bibnamefont {Rossmeisl}},\ }\href
  {\doibase 10.1021/acscatal.9b04343} {\bibfield  {journal} {\bibinfo
  {journal} {ACS Catalysis}\ }\textbf {\bibinfo {volume} {10}},\ \bibinfo
  {pages} {2169} (\bibinfo {year} {2020})},\ \Eprint
  {http://arxiv.org/abs/https://doi.org/10.1021/acscatal.9b04343}
  {https://doi.org/10.1021/acscatal.9b04343} \BibitemShut {NoStop}%
\bibitem [{\citenamefont {Jien-Wei}(2006)}]{Yeh2006}%
  \BibitemOpen
  \bibfield  {author} {\bibinfo {author} {\bibfnamefont {Y.}~\bibnamefont
  {Jien-Wei}},\ }\href@noop {} {\bibfield  {journal} {\bibinfo  {journal} {Ann.
  Chim. Sci. Mat}\ }\textbf {\bibinfo {volume} {31}},\ \bibinfo {pages} {633}
  (\bibinfo {year} {2006})}\BibitemShut {NoStop}%
\bibitem [{\citenamefont {Wang}\ \emph {et~al.}(2020)\citenamefont {Wang},
  \citenamefont {Tang}, \citenamefont {Li}, \citenamefont {Ai}, \citenamefont
  {Li}, \citenamefont {Xiao}, \citenamefont {Zhu}, \citenamefont {Liu},\ and\
  \citenamefont {Bai}}]{Wang2020}%
  \BibitemOpen
  \bibfield  {author} {\bibinfo {author} {\bibfnamefont {R.}~\bibnamefont
  {Wang}}, \bibinfo {author} {\bibfnamefont {Y.}~\bibnamefont {Tang}}, \bibinfo
  {author} {\bibfnamefont {S.}~\bibnamefont {Li}}, \bibinfo {author}
  {\bibfnamefont {Y.}~\bibnamefont {Ai}}, \bibinfo {author} {\bibfnamefont
  {Y.}~\bibnamefont {Li}}, \bibinfo {author} {\bibfnamefont {B.}~\bibnamefont
  {Xiao}}, \bibinfo {author} {\bibfnamefont {L.}~\bibnamefont {Zhu}}, \bibinfo
  {author} {\bibfnamefont {X.}~\bibnamefont {Liu}}, \ and\ \bibinfo {author}
  {\bibfnamefont {S.}~\bibnamefont {Bai}},\ }\href {\doibase
  https://doi.org/10.1016/j.jallcom.2020.154099} {\bibfield  {journal}
  {\bibinfo  {journal} {Journal of Alloys and Compounds}\ }\textbf {\bibinfo
  {volume} {825}},\ \bibinfo {pages} {154099} (\bibinfo {year}
  {2020})}\BibitemShut {NoStop}%
\bibitem [{\citenamefont {Yeh}(2013)}]{Yeh2013}%
  \BibitemOpen
  \bibfield  {author} {\bibinfo {author} {\bibfnamefont {J.-W.}\ \bibnamefont
  {Yeh}},\ }\href {\doibase 10.1007/s11837-013-0761-6} {\bibfield  {journal}
  {\bibinfo  {journal} {{JOM}}\ }\textbf {\bibinfo {volume} {65}},\ \bibinfo
  {pages} {1759} (\bibinfo {year} {2013})}\BibitemShut {NoStop}%
\bibitem [{\citenamefont {Pickering}\ and\ \citenamefont
  {Jones}(2016)}]{Pickering2016}%
  \BibitemOpen
  \bibfield  {author} {\bibinfo {author} {\bibfnamefont {E.~J.}\ \bibnamefont
  {Pickering}}\ and\ \bibinfo {author} {\bibfnamefont {N.~G.}\ \bibnamefont
  {Jones}},\ }\href {\doibase 10.1080/09506608.2016.1180020} {\bibfield
  {journal} {\bibinfo  {journal} {International Materials Reviews}\ }\textbf
  {\bibinfo {volume} {61}},\ \bibinfo {pages} {183} (\bibinfo {year}
  {2016})}\BibitemShut {NoStop}%
\bibitem [{\citenamefont {Miracle}\ and\ \citenamefont
  {Senkov}(2017)}]{Miracle2017}%
  \BibitemOpen
  \bibfield  {author} {\bibinfo {author} {\bibfnamefont {D.}~\bibnamefont
  {Miracle}}\ and\ \bibinfo {author} {\bibfnamefont {O.}~\bibnamefont
  {Senkov}},\ }\href {\doibase https://doi.org/10.1016/j.actamat.2016.08.081}
  {\bibfield  {journal} {\bibinfo  {journal} {Acta Materialia}\ }\textbf
  {\bibinfo {volume} {122}},\ \bibinfo {pages} {448} (\bibinfo {year}
  {2017})}\BibitemShut {NoStop}%
\bibitem [{\citenamefont {Szlachta}\ \emph {et~al.}(2014)\citenamefont
  {Szlachta}, \citenamefont {Bart{\'o}k},\ and\ \citenamefont
  {Cs{\'a}nyi}}]{szla+14prb}%
  \BibitemOpen
  \bibfield  {author} {\bibinfo {author} {\bibfnamefont {W.~J.}\ \bibnamefont
  {Szlachta}}, \bibinfo {author} {\bibfnamefont {A.~P.}\ \bibnamefont
  {Bart{\'o}k}}, \ and\ \bibinfo {author} {\bibfnamefont {G.}~\bibnamefont
  {Cs{\'a}nyi}},\ }\href {\doibase 10.1103/PhysRevB.90.104108} {\bibfield
  {journal} {\bibinfo  {journal} {Phys. Rev. B}\ }\textbf {\bibinfo {volume}
  {90}},\ \bibinfo {pages} {104108} (\bibinfo {year} {2014})}\BibitemShut
  {NoStop}%
\bibitem [{\citenamefont {Zuo}\ \emph {et~al.}(2020)\citenamefont {Zuo},
  \citenamefont {Chen}, \citenamefont {Li}, \citenamefont {Deng}, \citenamefont
  {Chen}, \citenamefont {Behler}, \citenamefont {Cs{\'a}nyi}, \citenamefont
  {Shapeev}, \citenamefont {Thompson}, \citenamefont {Wood},\ and\
  \citenamefont {Ong}}]{zuo+20jpcl}%
  \BibitemOpen
  \bibfield  {author} {\bibinfo {author} {\bibfnamefont {Y.}~\bibnamefont
  {Zuo}}, \bibinfo {author} {\bibfnamefont {C.}~\bibnamefont {Chen}}, \bibinfo
  {author} {\bibfnamefont {X.}~\bibnamefont {Li}}, \bibinfo {author}
  {\bibfnamefont {Z.}~\bibnamefont {Deng}}, \bibinfo {author} {\bibfnamefont
  {Y.}~\bibnamefont {Chen}}, \bibinfo {author} {\bibfnamefont {J.}~\bibnamefont
  {Behler}}, \bibinfo {author} {\bibfnamefont {G.}~\bibnamefont {Cs{\'a}nyi}},
  \bibinfo {author} {\bibfnamefont {A.~V.}\ \bibnamefont {Shapeev}}, \bibinfo
  {author} {\bibfnamefont {A.~P.}\ \bibnamefont {Thompson}}, \bibinfo {author}
  {\bibfnamefont {M.~A.}\ \bibnamefont {Wood}}, \ and\ \bibinfo {author}
  {\bibfnamefont {S.~P.}\ \bibnamefont {Ong}},\ }\href {\doibase
  10.1021/acs.jpca.9b08723} {\bibfield  {journal} {\bibinfo  {journal} {J.
  Phys. Chem. A}\ ,\ \bibinfo {pages} {acs.jpca.9b08723}} (\bibinfo {year}
  {2020})}\BibitemShut {NoStop}%
\bibitem [{\citenamefont {Lopanitsyna}\ \emph {et~al.}(2021)\citenamefont
  {Lopanitsyna}, \citenamefont {Ben~Mahmoud},\ and\ \citenamefont
  {Ceriotti}}]{lopa+21prm}%
  \BibitemOpen
  \bibfield  {author} {\bibinfo {author} {\bibfnamefont {N.}~\bibnamefont
  {Lopanitsyna}}, \bibinfo {author} {\bibfnamefont {C.}~\bibnamefont
  {Ben~Mahmoud}}, \ and\ \bibinfo {author} {\bibfnamefont {M.}~\bibnamefont
  {Ceriotti}},\ }\href {\doibase 10.1103/PhysRevMaterials.5.043802} {\bibfield
  {journal} {\bibinfo  {journal} {Phys. Rev. Materials}\ }\textbf {\bibinfo
  {volume} {5}},\ \bibinfo {pages} {043802} (\bibinfo {year}
  {2021})}\BibitemShut {NoStop}%
\bibitem [{\citenamefont {Ferrari}\ \emph {et~al.}(2020)\citenamefont
  {Ferrari}, \citenamefont {Dutta}, \citenamefont {Gubaev}, \citenamefont
  {Ikeda}, \citenamefont {Srinivasan}, \citenamefont {Grabowski},\ and\
  \citenamefont {K\"{o}rmann}}]{Ferrari2020a}%
  \BibitemOpen
  \bibfield  {author} {\bibinfo {author} {\bibfnamefont {A.}~\bibnamefont
  {Ferrari}}, \bibinfo {author} {\bibfnamefont {B.}~\bibnamefont {Dutta}},
  \bibinfo {author} {\bibfnamefont {K.}~\bibnamefont {Gubaev}}, \bibinfo
  {author} {\bibfnamefont {Y.}~\bibnamefont {Ikeda}}, \bibinfo {author}
  {\bibfnamefont {P.}~\bibnamefont {Srinivasan}}, \bibinfo {author}
  {\bibfnamefont {B.}~\bibnamefont {Grabowski}}, \ and\ \bibinfo {author}
  {\bibfnamefont {F.}~\bibnamefont {K\"{o}rmann}},\ }\href {\doibase
  10.1063/5.0025310} {\bibfield  {journal} {\bibinfo  {journal} {Journal of
  Applied Physics}\ }\textbf {\bibinfo {volume} {128}} (\bibinfo {year}
  {2020}),\ 10.1063/5.0025310}\BibitemShut {NoStop}%
\bibitem [{\citenamefont {Farkas}\ and\ \citenamefont
  {Caro}(2018{\natexlab{a}})}]{fark-caro18jmr}%
  \BibitemOpen
  \bibfield  {author} {\bibinfo {author} {\bibfnamefont {D.}~\bibnamefont
  {Farkas}}\ and\ \bibinfo {author} {\bibfnamefont {A.}~\bibnamefont {Caro}},\
  }\href {\doibase 10.1557/jmr.2018.245} {\bibfield  {journal} {\bibinfo
  {journal} {J. Mater. Res.}\ }\textbf {\bibinfo {volume} {33}},\ \bibinfo
  {pages} {3218} (\bibinfo {year} {2018}{\natexlab{a}})}\BibitemShut {NoStop}%
\bibitem [{\citenamefont {Daramola}\ \emph {et~al.}(2022)\citenamefont
  {Daramola}, \citenamefont {Bonny}, \citenamefont {Adjanor}, \citenamefont
  {Domain}, \citenamefont {Monnet},\ and\ \citenamefont
  {Fraczkiewicz}}]{dara+22cms}%
  \BibitemOpen
  \bibfield  {author} {\bibinfo {author} {\bibfnamefont {A.}~\bibnamefont
  {Daramola}}, \bibinfo {author} {\bibfnamefont {G.}~\bibnamefont {Bonny}},
  \bibinfo {author} {\bibfnamefont {G.}~\bibnamefont {Adjanor}}, \bibinfo
  {author} {\bibfnamefont {C.}~\bibnamefont {Domain}}, \bibinfo {author}
  {\bibfnamefont {G.}~\bibnamefont {Monnet}}, \ and\ \bibinfo {author}
  {\bibfnamefont {A.}~\bibnamefont {Fraczkiewicz}},\ }\href {\doibase
  10.1016/j.commatsci.2021.111165} {\bibfield  {journal} {\bibinfo  {journal}
  {Computational Materials Science}\ }\textbf {\bibinfo {volume} {203}},\
  \bibinfo {pages} {111165} (\bibinfo {year} {2022})}\BibitemShut {NoStop}%
\bibitem [{\citenamefont {Farkas}\ and\ \citenamefont
  {Caro}(2018{\natexlab{b}})}]{Farkas2018}%
  \BibitemOpen
  \bibfield  {author} {\bibinfo {author} {\bibfnamefont {D.}~\bibnamefont
  {Farkas}}\ and\ \bibinfo {author} {\bibfnamefont {A.}~\bibnamefont {Caro}},\
  }\href {\doibase 10.1557/jmr.2018.245} {\bibfield  {journal} {\bibinfo
  {journal} {Journal of Materials Research}\ }\textbf {\bibinfo {volume}
  {33}},\ \bibinfo {pages} {3218} (\bibinfo {year}
  {2018}{\natexlab{b}})}\BibitemShut {NoStop}%
\bibitem [{\citenamefont {Byggm\"{a}star}\ \emph {et~al.}(2021)\citenamefont
  {Byggm\"{a}star}, \citenamefont {Nordlund},\ and\ \citenamefont
  {Djurabekova}}]{Byggmstar2021}%
  \BibitemOpen
  \bibfield  {author} {\bibinfo {author} {\bibfnamefont {J.}~\bibnamefont
  {Byggm\"{a}star}}, \bibinfo {author} {\bibfnamefont {K.}~\bibnamefont
  {Nordlund}}, \ and\ \bibinfo {author} {\bibfnamefont {F.}~\bibnamefont
  {Djurabekova}},\ }\href {\doibase 10.1103/physrevb.104.104101} {\bibfield
  {journal} {\bibinfo  {journal} {Physical Review B}\ }\textbf {\bibinfo
  {volume} {104}} (\bibinfo {year} {2021}),\
  10.1103/physrevb.104.104101}\BibitemShut {NoStop}%
\bibitem [{\citenamefont {Rosenbrock}\ \emph {et~al.}(2021)\citenamefont
  {Rosenbrock}, \citenamefont {Gubaev}, \citenamefont {Shapeev}, \citenamefont
  {P{\'a}rtay}, \citenamefont {Bernstein}, \citenamefont {Cs{\'a}nyi},\ and\
  \citenamefont {Hart}}]{rose+21npjcm}%
  \BibitemOpen
  \bibfield  {author} {\bibinfo {author} {\bibfnamefont {C.~W.}\ \bibnamefont
  {Rosenbrock}}, \bibinfo {author} {\bibfnamefont {K.}~\bibnamefont {Gubaev}},
  \bibinfo {author} {\bibfnamefont {A.~V.}\ \bibnamefont {Shapeev}}, \bibinfo
  {author} {\bibfnamefont {L.~B.}\ \bibnamefont {P{\'a}rtay}}, \bibinfo
  {author} {\bibfnamefont {N.}~\bibnamefont {Bernstein}}, \bibinfo {author}
  {\bibfnamefont {G.}~\bibnamefont {Cs{\'a}nyi}}, \ and\ \bibinfo {author}
  {\bibfnamefont {G.~L.~W.}\ \bibnamefont {Hart}},\ }\href {\doibase
  10.1038/s41524-020-00477-2} {\bibfield  {journal} {\bibinfo  {journal} {npj
  Comput Mater}\ }\textbf {\bibinfo {volume} {7}},\ \bibinfo {pages} {24}
  (\bibinfo {year} {2021})}\BibitemShut {NoStop}%
\bibitem [{\citenamefont {Zhou}\ \emph {et~al.}(2022)\citenamefont {Zhou},
  \citenamefont {Srinivasan}, \citenamefont {K{\"o}rmann}, \citenamefont
  {Grabowski}, \citenamefont {Smith}, \citenamefont {Goddard},\ and\
  \citenamefont {Duff}}]{zhou+22prb}%
  \BibitemOpen
  \bibfield  {author} {\bibinfo {author} {\bibfnamefont {Y.}~\bibnamefont
  {Zhou}}, \bibinfo {author} {\bibfnamefont {P.}~\bibnamefont {Srinivasan}},
  \bibinfo {author} {\bibfnamefont {F.}~\bibnamefont {K{\"o}rmann}}, \bibinfo
  {author} {\bibfnamefont {B.}~\bibnamefont {Grabowski}}, \bibinfo {author}
  {\bibfnamefont {R.}~\bibnamefont {Smith}}, \bibinfo {author} {\bibfnamefont
  {P.}~\bibnamefont {Goddard}}, \ and\ \bibinfo {author} {\bibfnamefont
  {A.~I.}\ \bibnamefont {Duff}},\ }\href {\doibase 10.1103/PhysRevB.105.214302}
  {\bibfield  {journal} {\bibinfo  {journal} {Phys. Rev. B}\ }\textbf {\bibinfo
  {volume} {105}},\ \bibinfo {pages} {214302} (\bibinfo {year}
  {2022})}\BibitemShut {NoStop}%
\bibitem [{\citenamefont {Behler}\ and\ \citenamefont
  {Parrinello}(2007)}]{behl-parr07prl}%
  \BibitemOpen
  \bibfield  {author} {\bibinfo {author} {\bibfnamefont {J.}~\bibnamefont
  {Behler}}\ and\ \bibinfo {author} {\bibfnamefont {M.}~\bibnamefont
  {Parrinello}},\ }\href {\doibase 10.1103/PhysRevLett.98.146401} {\bibfield
  {journal} {\bibinfo  {journal} {Phys. Rev. Lett.}\ }\textbf {\bibinfo
  {volume} {98}},\ \bibinfo {pages} {146401} (\bibinfo {year}
  {2007})}\BibitemShut {NoStop}%
\bibitem [{\citenamefont {Bart{\'o}k}\ \emph {et~al.}(2010)\citenamefont
  {Bart{\'o}k}, \citenamefont {Payne}, \citenamefont {Kondor},\ and\
  \citenamefont {Cs{\'a}nyi}}]{bart+10prl}%
  \BibitemOpen
  \bibfield  {author} {\bibinfo {author} {\bibfnamefont {A.~P.}\ \bibnamefont
  {Bart{\'o}k}}, \bibinfo {author} {\bibfnamefont {M.~C.}\ \bibnamefont
  {Payne}}, \bibinfo {author} {\bibfnamefont {R.}~\bibnamefont {Kondor}}, \
  and\ \bibinfo {author} {\bibfnamefont {G.}~\bibnamefont {Cs{\'a}nyi}},\
  }\href {\doibase 10.1103/PhysRevLett.104.136403} {\bibfield  {journal}
  {\bibinfo  {journal} {Phys. Rev. Lett.}\ }\textbf {\bibinfo {volume} {104}},\
  \bibinfo {pages} {136403} (\bibinfo {year} {2010})}\BibitemShut {NoStop}%
\bibitem [{\citenamefont {Chen}\ \emph {et~al.}(2019)\citenamefont {Chen},
  \citenamefont {Ye}, \citenamefont {Zuo}, \citenamefont {Zheng},\ and\
  \citenamefont {Ong}}]{chen+19cm}%
  \BibitemOpen
  \bibfield  {author} {\bibinfo {author} {\bibfnamefont {C.}~\bibnamefont
  {Chen}}, \bibinfo {author} {\bibfnamefont {W.}~\bibnamefont {Ye}}, \bibinfo
  {author} {\bibfnamefont {Y.}~\bibnamefont {Zuo}}, \bibinfo {author}
  {\bibfnamefont {C.}~\bibnamefont {Zheng}}, \ and\ \bibinfo {author}
  {\bibfnamefont {S.~P.}\ \bibnamefont {Ong}},\ }\href {\doibase
  10.1021/acs.chemmater.9b01294} {\bibfield  {journal} {\bibinfo  {journal}
  {Chem. Mater.}\ }\textbf {\bibinfo {volume} {31}},\ \bibinfo {pages} {3564}
  (\bibinfo {year} {2019})}\BibitemShut {NoStop}%
\bibitem [{\citenamefont {Chen}\ and\ \citenamefont {Ong}(2022)}]{Chen2022}%
  \BibitemOpen
  \bibfield  {author} {\bibinfo {author} {\bibfnamefont {C.}~\bibnamefont
  {Chen}}\ and\ \bibinfo {author} {\bibfnamefont {S.~P.}\ \bibnamefont {Ong}},\
  }\href {\doibase 10.48550/ARXIV.2202.02450} {\enquote {\bibinfo {title} {A
  universal graph deep learning interatomic potential for the periodic
  table},}\ } (\bibinfo {year} {2022})\BibitemShut {NoStop}%
\bibitem [{\citenamefont {Batzner}\ \emph {et~al.}(2022)\citenamefont
  {Batzner}, \citenamefont {Musaelian}, \citenamefont {Sun}, \citenamefont
  {Geiger}, \citenamefont {Mailoa}, \citenamefont {Kornbluth}, \citenamefont
  {Molinari}, \citenamefont {Smidt},\ and\ \citenamefont
  {Kozinsky}}]{bazt+22ncomm}%
  \BibitemOpen
  \bibfield  {author} {\bibinfo {author} {\bibfnamefont {S.}~\bibnamefont
  {Batzner}}, \bibinfo {author} {\bibfnamefont {A.}~\bibnamefont {Musaelian}},
  \bibinfo {author} {\bibfnamefont {L.}~\bibnamefont {Sun}}, \bibinfo {author}
  {\bibfnamefont {M.}~\bibnamefont {Geiger}}, \bibinfo {author} {\bibfnamefont
  {J.~P.}\ \bibnamefont {Mailoa}}, \bibinfo {author} {\bibfnamefont
  {M.}~\bibnamefont {Kornbluth}}, \bibinfo {author} {\bibfnamefont
  {N.}~\bibnamefont {Molinari}}, \bibinfo {author} {\bibfnamefont {T.~E.}\
  \bibnamefont {Smidt}}, \ and\ \bibinfo {author} {\bibfnamefont
  {B.}~\bibnamefont {Kozinsky}},\ }\href {\doibase 10.1038/s41467-022-29939-5}
  {\bibfield  {journal} {\bibinfo  {journal} {Nat Commun}\ }\textbf {\bibinfo
  {volume} {13}},\ \bibinfo {pages} {2453} (\bibinfo {year}
  {2022})}\BibitemShut {NoStop}%
\bibitem [{\citenamefont {Batatia}\ \emph {et~al.}(2022)\citenamefont
  {Batatia}, \citenamefont {Kovacs}, \citenamefont {Simm}, \citenamefont
  {Ortner},\ and\ \citenamefont {Csanyi}}]{bata+22nips}%
  \BibitemOpen
  \bibfield  {author} {\bibinfo {author} {\bibfnamefont {I.}~\bibnamefont
  {Batatia}}, \bibinfo {author} {\bibfnamefont {D.~P.}\ \bibnamefont {Kovacs}},
  \bibinfo {author} {\bibfnamefont {G.~N.~C.}\ \bibnamefont {Simm}}, \bibinfo
  {author} {\bibfnamefont {C.}~\bibnamefont {Ortner}}, \ and\ \bibinfo {author}
  {\bibfnamefont {G.}~\bibnamefont {Csanyi}},\ }in\ \href@noop {} {\emph
  {\bibinfo {booktitle} {Adv. {{Neural Inf}}. {{Process}}. {{Syst}}.}}},\
  \bibinfo {editor} {edited by\ \bibinfo {editor} {\bibfnamefont {A.~H.}\
  \bibnamefont {Oh}}, \bibinfo {editor} {\bibfnamefont {A.}~\bibnamefont
  {Agarwal}}, \bibinfo {editor} {\bibfnamefont {D.}~\bibnamefont {Belgrave}}, \
  and\ \bibinfo {editor} {\bibfnamefont {K.}~\bibnamefont {Cho}}}\ (\bibinfo
  {year} {2022})\BibitemShut {NoStop}%
\bibitem [{\citenamefont {Lopanitsyna}\ \emph {et~al.}(2023)\citenamefont
  {Lopanitsyna}, \citenamefont {Fraux}, \citenamefont {Springer}, \citenamefont
  {De},\ and\ \citenamefont {Ceriotti}}]{lopa+23prm}%
  \BibitemOpen
  \bibfield  {author} {\bibinfo {author} {\bibfnamefont {N.}~\bibnamefont
  {Lopanitsyna}}, \bibinfo {author} {\bibfnamefont {G.}~\bibnamefont {Fraux}},
  \bibinfo {author} {\bibfnamefont {M.~A.}\ \bibnamefont {Springer}}, \bibinfo
  {author} {\bibfnamefont {S.}~\bibnamefont {De}}, \ and\ \bibinfo {author}
  {\bibfnamefont {M.}~\bibnamefont {Ceriotti}},\ }\href {\doibase
  10.1103/PhysRevMaterials.7.045802} {\bibfield  {journal} {\bibinfo  {journal}
  {Phys. Rev. Materials}\ }\textbf {\bibinfo {volume} {7}},\ \bibinfo {pages}
  {045802} (\bibinfo {year} {2023})}\BibitemShut {NoStop}%
\bibitem [{\citenamefont {Willatt}\ \emph {et~al.}(2018)\citenamefont
  {Willatt}, \citenamefont {Musil},\ and\ \citenamefont
  {Ceriotti}}]{will+18pccp}%
  \BibitemOpen
  \bibfield  {author} {\bibinfo {author} {\bibfnamefont {M.~J.}\ \bibnamefont
  {Willatt}}, \bibinfo {author} {\bibfnamefont {F.}~\bibnamefont {Musil}}, \
  and\ \bibinfo {author} {\bibfnamefont {M.}~\bibnamefont {Ceriotti}},\ }\href
  {\doibase 10.1039/c8cp05921g} {\bibfield  {journal} {\bibinfo  {journal}
  {Phys. Chem. Chem. Phys.}\ }\textbf {\bibinfo {volume} {20}},\ \bibinfo
  {pages} {29661} (\bibinfo {year} {2018})}\BibitemShut {NoStop}%
\bibitem [{\citenamefont {Owen}\ \emph {et~al.}(2023)\citenamefont {Owen},
  \citenamefont {Torrisi}, \citenamefont {Xie}, \citenamefont {Batzner},
  \citenamefont {Coulter}, \citenamefont {Musaelian}, \citenamefont {Sun},\
  and\ \citenamefont {Kozinsky}}]{owen2023complexity}%
  \BibitemOpen
  \bibfield  {author} {\bibinfo {author} {\bibfnamefont {C.~J.}\ \bibnamefont
  {Owen}}, \bibinfo {author} {\bibfnamefont {S.~B.}\ \bibnamefont {Torrisi}},
  \bibinfo {author} {\bibfnamefont {Y.}~\bibnamefont {Xie}}, \bibinfo {author}
  {\bibfnamefont {S.}~\bibnamefont {Batzner}}, \bibinfo {author} {\bibfnamefont
  {J.}~\bibnamefont {Coulter}}, \bibinfo {author} {\bibfnamefont
  {A.}~\bibnamefont {Musaelian}}, \bibinfo {author} {\bibfnamefont
  {L.}~\bibnamefont {Sun}}, \ and\ \bibinfo {author} {\bibfnamefont
  {B.}~\bibnamefont {Kozinsky}},\ }\href@noop {} {\bibfield  {journal}
  {\bibinfo  {journal} {arXiv preprint arXiv:2302.12993}\ } (\bibinfo {year}
  {2023})}\BibitemShut {NoStop}%
\bibitem [{\citenamefont {Kresse}\ and\ \citenamefont
  {Furthm{\"u}ller}(1996)}]{VASP}%
  \BibitemOpen
  \bibfield  {author} {\bibinfo {author} {\bibfnamefont {G.}~\bibnamefont
  {Kresse}}\ and\ \bibinfo {author} {\bibfnamefont {J.}~\bibnamefont
  {Furthm{\"u}ller}},\ }\href@noop {} {\bibfield  {journal} {\bibinfo
  {journal} {Phys. Rev. B}\ }\textbf {\bibinfo {volume} {54}},\ \bibinfo
  {pages} {11169} (\bibinfo {year} {1996})}\BibitemShut {NoStop}%
\bibitem [{\citenamefont {Monkhorst}\ and\ \citenamefont
  {Pack}(1976)}]{monk-pack76prb}%
  \BibitemOpen
  \bibfield  {author} {\bibinfo {author} {\bibfnamefont {H.~J.}\ \bibnamefont
  {Monkhorst}}\ and\ \bibinfo {author} {\bibfnamefont {J.~D.}\ \bibnamefont
  {Pack}},\ }\href {\doibase 10.1103/PhysRevB.13.5188} {\bibfield  {journal}
  {\bibinfo  {journal} {Phys. Rev. B}\ }\textbf {\bibinfo {volume} {13}},\
  \bibinfo {pages} {5188} (\bibinfo {year} {1976})}\BibitemShut {NoStop}%
\bibitem [{\citenamefont {Kresse}\ and\ \citenamefont
  {Joubert}(1999)}]{kresse1999ultrasoft}%
  \BibitemOpen
  \bibfield  {author} {\bibinfo {author} {\bibfnamefont {G.}~\bibnamefont
  {Kresse}}\ and\ \bibinfo {author} {\bibfnamefont {D.}~\bibnamefont
  {Joubert}},\ }\href@noop {} {\bibfield  {journal} {\bibinfo  {journal}
  {Physical review b}\ }\textbf {\bibinfo {volume} {59}},\ \bibinfo {pages}
  {1758} (\bibinfo {year} {1999})}\BibitemShut {NoStop}%
\bibitem [{\citenamefont {Csonka}\ \emph {et~al.}(2009)\citenamefont {Csonka},
  \citenamefont {Perdew}, \citenamefont {Ruzsinszky}, \citenamefont
  {Philipsen}, \citenamefont {Leb{\`e}gue}, \citenamefont {Paier},
  \citenamefont {Vydrov},\ and\ \citenamefont
  {{\'A}ngy{\'a}n}}]{csonka2009assessing}%
  \BibitemOpen
  \bibfield  {author} {\bibinfo {author} {\bibfnamefont {G.~I.}\ \bibnamefont
  {Csonka}}, \bibinfo {author} {\bibfnamefont {J.~P.}\ \bibnamefont {Perdew}},
  \bibinfo {author} {\bibfnamefont {A.}~\bibnamefont {Ruzsinszky}}, \bibinfo
  {author} {\bibfnamefont {P.~H.}\ \bibnamefont {Philipsen}}, \bibinfo {author}
  {\bibfnamefont {S.}~\bibnamefont {Leb{\`e}gue}}, \bibinfo {author}
  {\bibfnamefont {J.}~\bibnamefont {Paier}}, \bibinfo {author} {\bibfnamefont
  {O.~A.}\ \bibnamefont {Vydrov}}, \ and\ \bibinfo {author} {\bibfnamefont
  {J.~G.}\ \bibnamefont {{\'A}ngy{\'a}n}},\ }\href@noop {} {\bibfield
  {journal} {\bibinfo  {journal} {Physical Review B}\ }\textbf {\bibinfo
  {volume} {79}},\ \bibinfo {pages} {155107} (\bibinfo {year}
  {2009})}\BibitemShut {NoStop}%
\bibitem [{\citenamefont {Chen}\ and\ \citenamefont
  {Millis}(2016)}]{Chen2016Spin}%
  \BibitemOpen
  \bibfield  {author} {\bibinfo {author} {\bibfnamefont {H.}~\bibnamefont
  {Chen}}\ and\ \bibinfo {author} {\bibfnamefont {A.~J.}\ \bibnamefont
  {Millis}},\ }\href {\doibase 10.1103/PhysRevB.93.045133} {\bibfield
  {journal} {\bibinfo  {journal} {Phys. Rev. B}\ }\textbf {\bibinfo {volume}
  {93}},\ \bibinfo {pages} {045133} (\bibinfo {year} {2016})}\BibitemShut
  {NoStop}%
\bibitem [{\citenamefont {Huebsch}\ \emph {et~al.}(2021)\citenamefont
  {Huebsch}, \citenamefont {Nomoto}, \citenamefont {Suzuki},\ and\
  \citenamefont {Arita}}]{Huebsch2021Benchmark}%
  \BibitemOpen
  \bibfield  {author} {\bibinfo {author} {\bibfnamefont {M.-T.}\ \bibnamefont
  {Huebsch}}, \bibinfo {author} {\bibfnamefont {T.}~\bibnamefont {Nomoto}},
  \bibinfo {author} {\bibfnamefont {M.-T.}\ \bibnamefont {Suzuki}}, \ and\
  \bibinfo {author} {\bibfnamefont {R.}~\bibnamefont {Arita}},\ }\href
  {\doibase 10.1103/PhysRevX.11.011031} {\bibfield  {journal} {\bibinfo
  {journal} {Phys. Rev. X}\ }\textbf {\bibinfo {volume} {11}},\ \bibinfo
  {pages} {011031} (\bibinfo {year} {2021})}\BibitemShut {NoStop}%
\bibitem [{\citenamefont {Dragoni}\ \emph {et~al.}(2018)\citenamefont
  {Dragoni}, \citenamefont {Daff}, \citenamefont {Cs\'anyi},\ and\
  \citenamefont {Marzari}}]{Dragoni2018Achieving}%
  \BibitemOpen
  \bibfield  {author} {\bibinfo {author} {\bibfnamefont {D.}~\bibnamefont
  {Dragoni}}, \bibinfo {author} {\bibfnamefont {T.~D.}\ \bibnamefont {Daff}},
  \bibinfo {author} {\bibfnamefont {G.}~\bibnamefont {Cs\'anyi}}, \ and\
  \bibinfo {author} {\bibfnamefont {N.}~\bibnamefont {Marzari}},\ }\href
  {\doibase 10.1103/PhysRevMaterials.2.013808} {\bibfield  {journal} {\bibinfo
  {journal} {Phys. Rev. Mater.}\ }\textbf {\bibinfo {volume} {2}},\ \bibinfo
  {pages} {013808} (\bibinfo {year} {2018})}\BibitemShut {NoStop}%
\bibitem [{\citenamefont {Eckhoff}\ and\ \citenamefont
  {Behler}(2021)}]{eckh-behl21npjcm}%
  \BibitemOpen
  \bibfield  {author} {\bibinfo {author} {\bibfnamefont {M.}~\bibnamefont
  {Eckhoff}}\ and\ \bibinfo {author} {\bibfnamefont {J.}~\bibnamefont
  {Behler}},\ }\href {\doibase 10.1038/s41524-021-00636-z} {\bibfield
  {journal} {\bibinfo  {journal} {npj Comput Mater}\ }\textbf {\bibinfo
  {volume} {7}},\ \bibinfo {pages} {170} (\bibinfo {year} {2021})}\BibitemShut
  {NoStop}%
\bibitem [{\citenamefont {Novikov}\ \emph {et~al.}(2022)\citenamefont
  {Novikov}, \citenamefont {Grabowski}, \citenamefont {K{\"o}rmann},\ and\
  \citenamefont {Shapeev}}]{Novikov2022Magnetic}%
  \BibitemOpen
  \bibfield  {author} {\bibinfo {author} {\bibfnamefont {I.}~\bibnamefont
  {Novikov}}, \bibinfo {author} {\bibfnamefont {B.}~\bibnamefont {Grabowski}},
  \bibinfo {author} {\bibfnamefont {F.}~\bibnamefont {K{\"o}rmann}}, \ and\
  \bibinfo {author} {\bibfnamefont {A.}~\bibnamefont {Shapeev}},\ }\href@noop
  {} {\bibfield  {journal} {\bibinfo  {journal} {npj Computational Materials}\
  }\textbf {\bibinfo {volume} {8}},\ \bibinfo {pages} {13} (\bibinfo {year}
  {2022})}\BibitemShut {NoStop}%
\bibitem [{\citenamefont {Drautz}(2020)}]{drau20prb}%
  \BibitemOpen
  \bibfield  {author} {\bibinfo {author} {\bibfnamefont {R.}~\bibnamefont
  {Drautz}},\ }\href {\doibase 10.1103/PhysRevB.102.024104} {\bibfield
  {journal} {\bibinfo  {journal} {Phys. Rev. B}\ }\textbf {\bibinfo {volume}
  {102}},\ \bibinfo {pages} {024104} (\bibinfo {year} {2020})}\BibitemShut
  {NoStop}%
\bibitem [{\citenamefont {Domina}\ \emph {et~al.}(2022)\citenamefont {Domina},
  \citenamefont {Cobelli},\ and\ \citenamefont {Sanvito}}]{domi+22prb}%
  \BibitemOpen
  \bibfield  {author} {\bibinfo {author} {\bibfnamefont {M.}~\bibnamefont
  {Domina}}, \bibinfo {author} {\bibfnamefont {M.}~\bibnamefont {Cobelli}}, \
  and\ \bibinfo {author} {\bibfnamefont {S.}~\bibnamefont {Sanvito}},\ }\href
  {\doibase 10.1103/PhysRevB.105.214439} {\bibfield  {journal} {\bibinfo
  {journal} {Phys. Rev. B}\ }\textbf {\bibinfo {volume} {105}},\ \bibinfo
  {pages} {214439} (\bibinfo {year} {2022})}\BibitemShut {NoStop}%
\bibitem [{\citenamefont {Imbalzano}\ \emph {et~al.}(2018)\citenamefont
  {Imbalzano}, \citenamefont {Anelli}, \citenamefont {Giofr{\'e}},
  \citenamefont {Klees}, \citenamefont {Behler},\ and\ \citenamefont
  {Ceriotti}}]{imba+18jcp}%
  \BibitemOpen
  \bibfield  {author} {\bibinfo {author} {\bibfnamefont {G.}~\bibnamefont
  {Imbalzano}}, \bibinfo {author} {\bibfnamefont {A.}~\bibnamefont {Anelli}},
  \bibinfo {author} {\bibfnamefont {D.}~\bibnamefont {Giofr{\'e}}}, \bibinfo
  {author} {\bibfnamefont {S.}~\bibnamefont {Klees}}, \bibinfo {author}
  {\bibfnamefont {J.}~\bibnamefont {Behler}}, \ and\ \bibinfo {author}
  {\bibfnamefont {M.}~\bibnamefont {Ceriotti}},\ }\href {\doibase
  10.1063/1.5024611} {\bibfield  {journal} {\bibinfo  {journal} {J. Chem.
  Phys.}\ }\textbf {\bibinfo {volume} {148}},\ \bibinfo {pages} {241730}
  (\bibinfo {year} {2018})},\ \Eprint {http://arxiv.org/abs/1804.02150}
  {arxiv:1804.02150} \BibitemShut {NoStop}%
\bibitem [{\citenamefont {Pozdnyakov}\ and\ \citenamefont
  {Ceriotti}(2023)}]{pozdnyakov2023smooth}%
  \BibitemOpen
  \bibfield  {author} {\bibinfo {author} {\bibfnamefont {S.~N.}\ \bibnamefont
  {Pozdnyakov}}\ and\ \bibinfo {author} {\bibfnamefont {M.}~\bibnamefont
  {Ceriotti}},\ }\href@noop {} {\bibfield  {journal} {\bibinfo  {journal}
  {arXiv preprint arXiv:2305.19302}\ } (\bibinfo {year} {2023})}\BibitemShut
  {NoStop}%
\bibitem [{\citenamefont {Mazitov}\ \emph {et~al.}(2023)\citenamefont
  {Mazitov}, \citenamefont {Springer}, \citenamefont {Lopanitsyna},
  \citenamefont {Fraux}, \citenamefont {De},\ and\ \citenamefont
  {Ceriotti}}]{hea25sdataset}%
  \BibitemOpen
  \bibfield  {author} {\bibinfo {author} {\bibfnamefont {A.}~\bibnamefont
  {Mazitov}}, \bibinfo {author} {\bibfnamefont {M.~A.}\ \bibnamefont
  {Springer}}, \bibinfo {author} {\bibfnamefont {N.}~\bibnamefont
  {Lopanitsyna}}, \bibinfo {author} {\bibfnamefont {G.}~\bibnamefont {Fraux}},
  \bibinfo {author} {\bibfnamefont {S.}~\bibnamefont {De}}, \ and\ \bibinfo
  {author} {\bibfnamefont {M.}~\bibnamefont {Ceriotti}},\ }\href {\doibase
  10.24435/materialscloud:ps-20} {\enquote {\bibinfo {title} {Surface
  segregation in high-entropy alloys from alchemical machine learning: dataset
  hea25s},}\ }\bibinfo {howpublished}
  {\url{https://doi.org/10.24435/materialscloud:ps-20}} (\bibinfo {year}
  {2023})\BibitemShut {NoStop}%
\bibitem [{\citenamefont {Ferrari}\ and\ \citenamefont
  {K\"{o}rmann}(2020)}]{Ferrari2020}%
  \BibitemOpen
  \bibfield  {author} {\bibinfo {author} {\bibfnamefont {A.}~\bibnamefont
  {Ferrari}}\ and\ \bibinfo {author} {\bibfnamefont {F.}~\bibnamefont
  {K\"{o}rmann}},\ }\href {\doibase 10.1016/j.apsusc.2020.147471} {\bibfield
  {journal} {\bibinfo  {journal} {Applied Surface Science}\ }\textbf {\bibinfo
  {volume} {533}},\ \bibinfo {pages} {147471} (\bibinfo {year}
  {2020})}\BibitemShut {NoStop}%
\bibitem [{\citenamefont {Wynblatt}\ and\ \citenamefont
  {Chatain}(2019)}]{Wynblatt2019}%
  \BibitemOpen
  \bibfield  {author} {\bibinfo {author} {\bibfnamefont {P.}~\bibnamefont
  {Wynblatt}}\ and\ \bibinfo {author} {\bibfnamefont {D.}~\bibnamefont
  {Chatain}},\ }\href {\doibase 10.1103/PhysRevMaterials.3.054004} {\bibfield
  {journal} {\bibinfo  {journal} {Phys. Rev. Mater.}\ }\textbf {\bibinfo
  {volume} {3}},\ \bibinfo {pages} {054004} (\bibinfo {year}
  {2019})}\BibitemShut {NoStop}%
\bibitem [{\citenamefont {Dahale}\ \emph {et~al.}(2022)\citenamefont {Dahale},
  \citenamefont {Goverapet~Srinivasan}, \citenamefont {Mishra}, \citenamefont
  {Maiti},\ and\ \citenamefont {Rai}}]{Dahale2022}%
  \BibitemOpen
  \bibfield  {author} {\bibinfo {author} {\bibfnamefont {C.}~\bibnamefont
  {Dahale}}, \bibinfo {author} {\bibfnamefont {S.}~\bibnamefont
  {Goverapet~Srinivasan}}, \bibinfo {author} {\bibfnamefont {S.}~\bibnamefont
  {Mishra}}, \bibinfo {author} {\bibfnamefont {S.}~\bibnamefont {Maiti}}, \
  and\ \bibinfo {author} {\bibfnamefont {B.}~\bibnamefont {Rai}},\ }\href
  {\doibase 10.1039/D2ME00045H} {\bibfield  {journal} {\bibinfo  {journal}
  {Mol. Syst. Des. Eng.}\ }\textbf {\bibinfo {volume} {7}},\ \bibinfo {pages}
  {878} (\bibinfo {year} {2022})}\BibitemShut {NoStop}%
\bibitem [{\citenamefont {Chatain}\ and\ \citenamefont
  {Wynblatt}(2021)}]{Chatain2021}%
  \BibitemOpen
  \bibfield  {author} {\bibinfo {author} {\bibfnamefont {D.}~\bibnamefont
  {Chatain}}\ and\ \bibinfo {author} {\bibfnamefont {P.}~\bibnamefont
  {Wynblatt}},\ }\href {\doibase
  https://doi.org/10.1016/j.commatsci.2020.110101} {\bibfield  {journal}
  {\bibinfo  {journal} {Computational Materials Science}\ }\textbf {\bibinfo
  {volume} {187}},\ \bibinfo {pages} {110101} (\bibinfo {year}
  {2021})}\BibitemShut {NoStop}%
\bibitem [{\citenamefont {Middleburgh}\ \emph {et~al.}(2014)\citenamefont
  {Middleburgh}, \citenamefont {King}, \citenamefont {Lumpkin}, \citenamefont
  {Cortie},\ and\ \citenamefont {Edwards}}]{Middleburgh2014}%
  \BibitemOpen
  \bibfield  {author} {\bibinfo {author} {\bibfnamefont {S.}~\bibnamefont
  {Middleburgh}}, \bibinfo {author} {\bibfnamefont {D.}~\bibnamefont {King}},
  \bibinfo {author} {\bibfnamefont {G.}~\bibnamefont {Lumpkin}}, \bibinfo
  {author} {\bibfnamefont {M.}~\bibnamefont {Cortie}}, \ and\ \bibinfo {author}
  {\bibfnamefont {L.}~\bibnamefont {Edwards}},\ }\href {\doibase
  https://doi.org/10.1016/j.jallcom.2014.01.135} {\bibfield  {journal}
  {\bibinfo  {journal} {Journal of Alloys and Compounds}\ }\textbf {\bibinfo
  {volume} {599}},\ \bibinfo {pages} {179} (\bibinfo {year}
  {2014})}\BibitemShut {NoStop}%
\bibitem [{\citenamefont {Guisbiers}\ \emph {et~al.}(2016)\citenamefont
  {Guisbiers}, \citenamefont {Mendoza-Cruz}, \citenamefont {Bazán-Díaz},
  \citenamefont {Velázquez-Salazar}, \citenamefont {Mendoza-Perez},
  \citenamefont {Robledo-Torres}, \citenamefont {Rodriguez-Lopez},
  \citenamefont {Montejano-Carrizales}, \citenamefont {Whetten},\ and\
  \citenamefont {José-Yacamán}}]{Guisbiers2016}%
  \BibitemOpen
  \bibfield  {author} {\bibinfo {author} {\bibfnamefont {G.}~\bibnamefont
  {Guisbiers}}, \bibinfo {author} {\bibfnamefont {R.}~\bibnamefont
  {Mendoza-Cruz}}, \bibinfo {author} {\bibfnamefont {L.}~\bibnamefont
  {Bazán-Díaz}}, \bibinfo {author} {\bibfnamefont {J.~J.}\ \bibnamefont
  {Velázquez-Salazar}}, \bibinfo {author} {\bibfnamefont {R.}~\bibnamefont
  {Mendoza-Perez}}, \bibinfo {author} {\bibfnamefont {J.~A.}\ \bibnamefont
  {Robledo-Torres}}, \bibinfo {author} {\bibfnamefont {J.-L.}\ \bibnamefont
  {Rodriguez-Lopez}}, \bibinfo {author} {\bibfnamefont {J.~M.}\ \bibnamefont
  {Montejano-Carrizales}}, \bibinfo {author} {\bibfnamefont {R.~L.}\
  \bibnamefont {Whetten}}, \ and\ \bibinfo {author} {\bibfnamefont
  {M.}~\bibnamefont {José-Yacamán}},\ }\href {\doibase
  10.1021/acsnano.5b05755} {\bibfield  {journal} {\bibinfo  {journal} {ACS
  Nano}\ }\textbf {\bibinfo {volume} {10}},\ \bibinfo {pages} {188} (\bibinfo
  {year} {2016})},\ \bibinfo {note} {pMID: 26605557},\ \Eprint
  {http://arxiv.org/abs/https://doi.org/10.1021/acsnano.5b05755}
  {https://doi.org/10.1021/acsnano.5b05755} \BibitemShut {NoStop}%
\bibitem [{\citenamefont {Pedersen}\ \emph
  {et~al.}(2021{\natexlab{b}})\citenamefont {Pedersen}, \citenamefont
  {Batchelor}, \citenamefont {Yan}, \citenamefont {Skjegstad},\ and\
  \citenamefont {Rossmeisl}}]{Pedersen2021-2}%
  \BibitemOpen
  \bibfield  {author} {\bibinfo {author} {\bibfnamefont {J.~K.}\ \bibnamefont
  {Pedersen}}, \bibinfo {author} {\bibfnamefont {T.~A.}\ \bibnamefont
  {Batchelor}}, \bibinfo {author} {\bibfnamefont {D.}~\bibnamefont {Yan}},
  \bibinfo {author} {\bibfnamefont {L.~E.~J.}\ \bibnamefont {Skjegstad}}, \
  and\ \bibinfo {author} {\bibfnamefont {J.}~\bibnamefont {Rossmeisl}},\ }\href
  {\doibase https://doi.org/10.1016/j.coelec.2020.100651} {\bibfield  {journal}
  {\bibinfo  {journal} {Current Opinion in Electrochemistry}\ }\textbf
  {\bibinfo {volume} {26}},\ \bibinfo {pages} {100651} (\bibinfo {year}
  {2021}{\natexlab{b}})}\BibitemShut {NoStop}%
\bibitem [{\citenamefont {Adam}(1941)}]{adam41book}%
  \BibitemOpen
  \bibfield  {author} {\bibinfo {author} {\bibfnamefont {N.~K.}\ \bibnamefont
  {Adam}},\ }\href@noop {} {\emph {\bibinfo {title} {The {{Physics}} and
  {{Chemistry}} of {{Surfaces}}}}}\ (\bibinfo  {publisher} {{Oxford University
  Press}},\ \bibinfo {year} {1941})\BibitemShut {NoStop}%
\bibitem [{\citenamefont {Shiihara}\ \emph {et~al.}(2013)\citenamefont
  {Shiihara}, \citenamefont {Kohyama},\ and\ \citenamefont
  {Ishibashi}}]{Shiihara2013Origin}%
  \BibitemOpen
  \bibfield  {author} {\bibinfo {author} {\bibfnamefont {Y.}~\bibnamefont
  {Shiihara}}, \bibinfo {author} {\bibfnamefont {M.}~\bibnamefont {Kohyama}}, \
  and\ \bibinfo {author} {\bibfnamefont {S.}~\bibnamefont {Ishibashi}},\ }\href
  {\doibase 10.1103/physrevb.87.125430} {\bibfield  {journal} {\bibinfo
  {journal} {Physical Review B}\ }\textbf {\bibinfo {volume} {87}} (\bibinfo
  {year} {2013}),\ 10.1103/physrevb.87.125430}\BibitemShut {NoStop}%
\bibitem [{\citenamefont {Shiihara}\ and\ \citenamefont
  {Kohyama}(2016)}]{Shiihara2016Contribution}%
  \BibitemOpen
  \bibfield  {author} {\bibinfo {author} {\bibfnamefont {Y.}~\bibnamefont
  {Shiihara}}\ and\ \bibinfo {author} {\bibfnamefont {M.}~\bibnamefont
  {Kohyama}},\ }\href {\doibase 10.1016/j.susc.2015.08.009} {\bibfield
  {journal} {\bibinfo  {journal} {Surface Science}\ }\textbf {\bibinfo {volume}
  {644}},\ \bibinfo {pages} {122–128} (\bibinfo {year} {2016})}\BibitemShut
  {NoStop}%
\bibitem [{\citenamefont {Maulana}\ \emph {et~al.}(2023)\citenamefont
  {Maulana}, \citenamefont {Chen}, \citenamefont {Shi}, \citenamefont {Yang},
  \citenamefont {Lizandara-Pueyo}, \citenamefont {Seeler}, \citenamefont
  {Abruña}, \citenamefont {Muller}, \citenamefont {Schierle-Arndt},\ and\
  \citenamefont {Yang}}]{Maulana2023}%
  \BibitemOpen
  \bibfield  {author} {\bibinfo {author} {\bibfnamefont {A.~L.}\ \bibnamefont
  {Maulana}}, \bibinfo {author} {\bibfnamefont {P.-C.}\ \bibnamefont {Chen}},
  \bibinfo {author} {\bibfnamefont {Z.}~\bibnamefont {Shi}}, \bibinfo {author}
  {\bibfnamefont {Y.}~\bibnamefont {Yang}}, \bibinfo {author} {\bibfnamefont
  {C.}~\bibnamefont {Lizandara-Pueyo}}, \bibinfo {author} {\bibfnamefont
  {F.}~\bibnamefont {Seeler}}, \bibinfo {author} {\bibfnamefont {H.~D.}\
  \bibnamefont {Abruña}}, \bibinfo {author} {\bibfnamefont {D.}~\bibnamefont
  {Muller}}, \bibinfo {author} {\bibfnamefont {K.}~\bibnamefont
  {Schierle-Arndt}}, \ and\ \bibinfo {author} {\bibfnamefont {P.}~\bibnamefont
  {Yang}},\ }\href {\doibase 10.1021/acs.nanolett.3c01831} {\bibfield
  {journal} {\bibinfo  {journal} {Nano Letters}\ }\textbf {\bibinfo {volume}
  {23}},\ \bibinfo {pages} {6637} (\bibinfo {year} {2023})},\ \bibinfo {note}
  {pMID: 37406363},\ \Eprint
  {http://arxiv.org/abs/https://doi.org/10.1021/acs.nanolett.3c01831}
  {https://doi.org/10.1021/acs.nanolett.3c01831} \BibitemShut {NoStop}%
\bibitem [{\citenamefont {Wang}\ \emph
  {et~al.}(2021{\natexlab{b}})\citenamefont {Wang}, \citenamefont {Wu},
  \citenamefont {Fu}, \citenamefont {Liu}, \citenamefont {Ren}, \citenamefont
  {Yan}, \citenamefont {Chen}, \citenamefont {Lan}, \citenamefont {Hahn},\ and\
  \citenamefont {Feng}}]{Wang2021Nanocrystalline}%
  \BibitemOpen
  \bibfield  {author} {\bibinfo {author} {\bibfnamefont {J.}~\bibnamefont
  {Wang}}, \bibinfo {author} {\bibfnamefont {S.}~\bibnamefont {Wu}}, \bibinfo
  {author} {\bibfnamefont {S.}~\bibnamefont {Fu}}, \bibinfo {author}
  {\bibfnamefont {S.}~\bibnamefont {Liu}}, \bibinfo {author} {\bibfnamefont
  {Z.}~\bibnamefont {Ren}}, \bibinfo {author} {\bibfnamefont {M.}~\bibnamefont
  {Yan}}, \bibinfo {author} {\bibfnamefont {S.}~\bibnamefont {Chen}}, \bibinfo
  {author} {\bibfnamefont {S.}~\bibnamefont {Lan}}, \bibinfo {author}
  {\bibfnamefont {H.}~\bibnamefont {Hahn}}, \ and\ \bibinfo {author}
  {\bibfnamefont {T.}~\bibnamefont {Feng}},\ }\href@noop {} {\bibfield
  {journal} {\bibinfo  {journal} {Journal of Materials Science \& Technology}\
  }\textbf {\bibinfo {volume} {77}},\ \bibinfo {pages} {126} (\bibinfo {year}
  {2021}{\natexlab{b}})}\BibitemShut {NoStop}%
\bibitem [{\citenamefont {Ming}\ \emph {et~al.}(2023)\citenamefont {Ming},
  \citenamefont {Qiang}, \citenamefont {Weiwu}, \citenamefont {Xiaoting},
  \citenamefont {Jia},\ and\ \citenamefont {Jian}}]{IrFeCo_patent}%
  \BibitemOpen
  \bibfield  {author} {\bibinfo {author} {\bibfnamefont {T.}~\bibnamefont
  {Ming}}, \bibinfo {author} {\bibfnamefont {C.}~\bibnamefont {Qiang}},
  \bibinfo {author} {\bibfnamefont {S.}~\bibnamefont {Weiwu}}, \bibinfo
  {author} {\bibfnamefont {L.}~\bibnamefont {Xiaoting}}, \bibinfo {author}
  {\bibfnamefont {W.}~\bibnamefont {Jia}}, \ and\ \bibinfo {author}
  {\bibfnamefont {X.}~\bibnamefont {Jian}},\ }\href
  {https://patents.google.com/patent/CN113215616B/en} {\enquote {\bibinfo
  {title} {{I}r{C}o{F}e at {MX}ene composite catalyst and preparation method
  and application thereof},}\ } (\bibinfo {year} {2023})\BibitemShut {NoStop}%
\end{thebibliography}
\end{document}